\documentclass[twocolumn]{aastex63}
\usepackage{amssymb,amsmath,amsopn}

\newcommand{\unit}[1]{\ensuremath{\, \mathrm{#1}}}
\newcommand{\figref}[1]{Figure \ref{#1}}

\newcommand{\code}[1]{\ensuremath{\textsf{#1}}}

\newcommand{\scipy}{\code{scipy}}
\newcommand{\fnc}[1]{\code{#1}}

\newcommand{\secref}[1]{$\S$\ref{#1}}

\newcommand{\halpha}{\ensuremath{H\alpha}}
\newcommand{\hfit}{\ensuremath{\halpha_\mathit{fit} }}
\newcommand{\linecenter}{\ensuremath{\lambda_0}}
\newcommand{\linew}{\ensuremath{\delta \lambda}}
\newcommand{\imin}{\ensuremath{I_\mathit{min}}}
\newcommand{\ihalf}{\ensuremath{I_\mathit{half}}}

\graphicspath{{./}{figures/}}

\received{January 21, 2022}
\revised{April 25, 2022}
\accepted{April 27, 2022}

\submitjournal{ApJS}

\shorttitle{The enhanced network from photosphere to corona}
\shortauthors{Kobelski et al.}

\begin{document}

\title{A publicly available multi-observatory data set of an enhanced network patch from the Photosphere to Corona}
\correspondingauthor{Adam R. Kobelski}
\email{adam.kobelski@nasa.gov}

\author[0000-0002-4691-1729]{Adam R. Kobelski}
\affiliation{NASA Marshall Space Flight Center}
\affiliation{West Virginia University}

\author[0000-0002-8259-8303]{Lucas A. Tarr}
\affiliation{National Solar Observatory}
\author[0000-0001-5459-2628]{Sarah A. Jaeggli}
\affiliation{National Solar Observatory}
\author{Nicholas Luber}
\affiliation{West Virginia University}
\author[0000-0001-6102-6851]{Harry P. Warren}
\affiliation{Naval Research Laboratory}
\author[0000-0002-6172-0517]{Sabrina Savage}
\affiliation{NASA Marshall Space Flight Center}

\begin{abstract}

New instruments sensitive to chromospheric radiation at X-ray, UV, Visible, IR, and sub-mm wavelengths have become available that significantly enhance our ability to understand the bi-directional flow of energy through the chromosphere.
We describe the calibration, co-alignment, initial results, and public release of a new data set combining a large number of these instruments to obtain multi-wavelength photospheric, chromospheric, and coronal observations capable of improving our understanding of the connectivity between the photosphere and the corona via transient brightenings and wave signatures.
The observations center on a bipolar region of enhanced network magnetic flux near disk center on SOL2017-03-17T14:00-17:00.
The comprehensive data set provides one of the most complete views of chromospheric activity related to small scale brightenings in the corona and chromosphere to date.
Our initial analysis shows strong spatial correspondence between the areas of broadest width of the Hydrogen-$\alpha$ spectral line and the hottest temperatures observed in ALMA Band 3 radio data, with a linear coefficient of $6.12\times 10^{-5}$\AA{}/K.
The correspondence persists for the duration of co-temporal observations ($\approx 60\unit{m}$).
Numerous transient brightenings were observed in multiple data series.
We highlight a single, well observed transient brightening along a set of thin filamentary features with a duration of 20 minutes.
The timing of the peak intensity transitions from the cooler (ALMA, 7000 K) to hotter (XRT, 3 MK) data series.
\end{abstract}

\section{Introduction} \label{sec:intro}
The methods of transporting energy from the photosphere through the chromosphere to the million degree corona have been debated since the discovery of the hot corona \citep[][]{1933LyotMarshall, 1939Grotrian}. 
The underlying problem is that thermal conduction, which is very efficient in solar plasmas, will transport energy from the hot corona to the cool chromosphere from which it rapidly radiates away, so the energy in the corona must constantly be resupplied. 
While many theories and mechanisms have been presented (e.g. magnetic waves, magnetic reconnection, and their interplay) to understand the flow of energy, a complete understanding remains elusive.  \citet{2015deMoortel} and \citet{2015Klimchuk_CHoverview} provide concise reviews of this long lasting, multifaceted problem.
One aspect that has eluded prior studies is a comprehensive insight into how wave motion through the chromosphere relates to the occurrence of small coronal flares, as well as how small flares drive chromospheric heating and waves. 

Observationally, one can follow plasma flows through the chromosphere and corona by using emission and absorption spectral measurements. 
For stable loops, raster scans from the \textit{Hinode} EUV Imaging Spectrograph (EIS) and the Interface Region Imaging Spectrograph (IRIS) have been used to better constrain the rate of heating in quiescent regions of the Sun \citep[e.g.][]{2016BrooksWarren_NTWidths,2016TestaEA_ARLoopsIRISEIS, 2017GhoshEA_LoopsIRISEIS}. 
Measurement of flows during transient events requires fortuitous pointing and specific observations with minimal rastering. 
Such studies are crucial to understanding the flow of energy into the corona, as well as providing statistical insight into this coupled system during solar flares.

In this paper, we present a unique and extensive data set with which to study the flow of energy through the chromosphere and the interplay between transients and wave activity.
Calibrated and co-aligned data in FITS format with WCS compliant coordinates are available through a publicly accessible archive hosted at the National Solar Observatory: \url{https://share.nso.edu/shared/dkist/ltarr/kolsch/}.
Even though the data set is of a quiescent solar region, it displays notable dynamics from the photosphere to the corona. 
This region is exemplary of the ubiquitous features found throughout the solar atmosphere during all phases of the solar cycle. 
Since this data set probes each layer of the Sun at multiple wavelengths, heights, and temperatures, it can be studied with distinct methodologies, and can provide multiple ground-truths and perspectives for which to compare to models of chromospheric energy flow and transfer in later papers.

The data set includes photospheric magnetic field information from the Helioseismic and Magnetic Imager \citep[HMI,][]{2012SchouEA_HMI} as well as imaging from the Atmospheric Imaging Assembly \citep[AIA,][]{2012LemenEA_AIA} on board the Solar Dynamics Observatory \citep[SDO,][]{2012PesnellEA_SDO}; photospheric magnetic field data from the Solar Optical Telescope \citep[SOT,][]{2008TsunetaEA_SOT}, coronal X-ray imaging from the X-Ray Telescope \citep[XRT,][]{2007GolubEA_XRT}, and extreme ultraviolet spectra from EUV Imaging Spectrometer \citep[EIS,][]{2007CulhaneEA_EIS} on board the {\it Hinode} spacecraft \citep[][]{2007KosugiEA_Hinode}; chromospheric images and spectra from the Interface Region Imaging Spectrograph \citep[IRIS,][]{2014DePontieuEA_IRIS}; spectral images from the Interferometric BIdimensional Spectrometer \citep[IBIS,][]{2006Cavallini_IBIS} and infrared spectropolarimetry from the Facility InfraRed Spectropolarimeter \citep[FIRS,][]{2010JaeggliEA_FIRS} at the Dunn Solar Telescope (DST); and millimeter wave information from the Atacama Large Millimeter Array \citep[ALMA,][]{2009WootenThompson_ALMA, 2010HillsEA_ALMA2}.  
For consistency, we refer to each individual instrument's collected data as a \emph{data series} and the full collection of data series as the \emph{data set}.

Our first task in this paper is to fully describe the observations from each data series, including all necessary reduction and calibration of the imaging and spectral data (\secref{sec:data}).
Next we describe the co-alignment process, which is nontrivial given the heterogeneous nature of the data set (\secref{sec:coalign}).
We then turn in \secref{sec:analysis} to two initial findings: we find a linear relation between ALMA Band 3 brightness temperature and the width of the hydrogen alpha spectral line, which extends a previous result found within active region plage into regions of weak network magnetic field that cover the bulk of the solar surface; and we identify a transient brightening that has clear signatures in multiple channels, sequentially extending from data series sensitive to chromospheric locations and processes, through transition region series, into coronal series, and then back.
We conclude in \secref{sec:conclusion}. 

\section{Data} \label{sec:data}
\subsection{Overview and Coordination}
\begin{figure}
    \centering
    \includegraphics[trim={0 10 0 0},clip,width=0.95\linewidth]{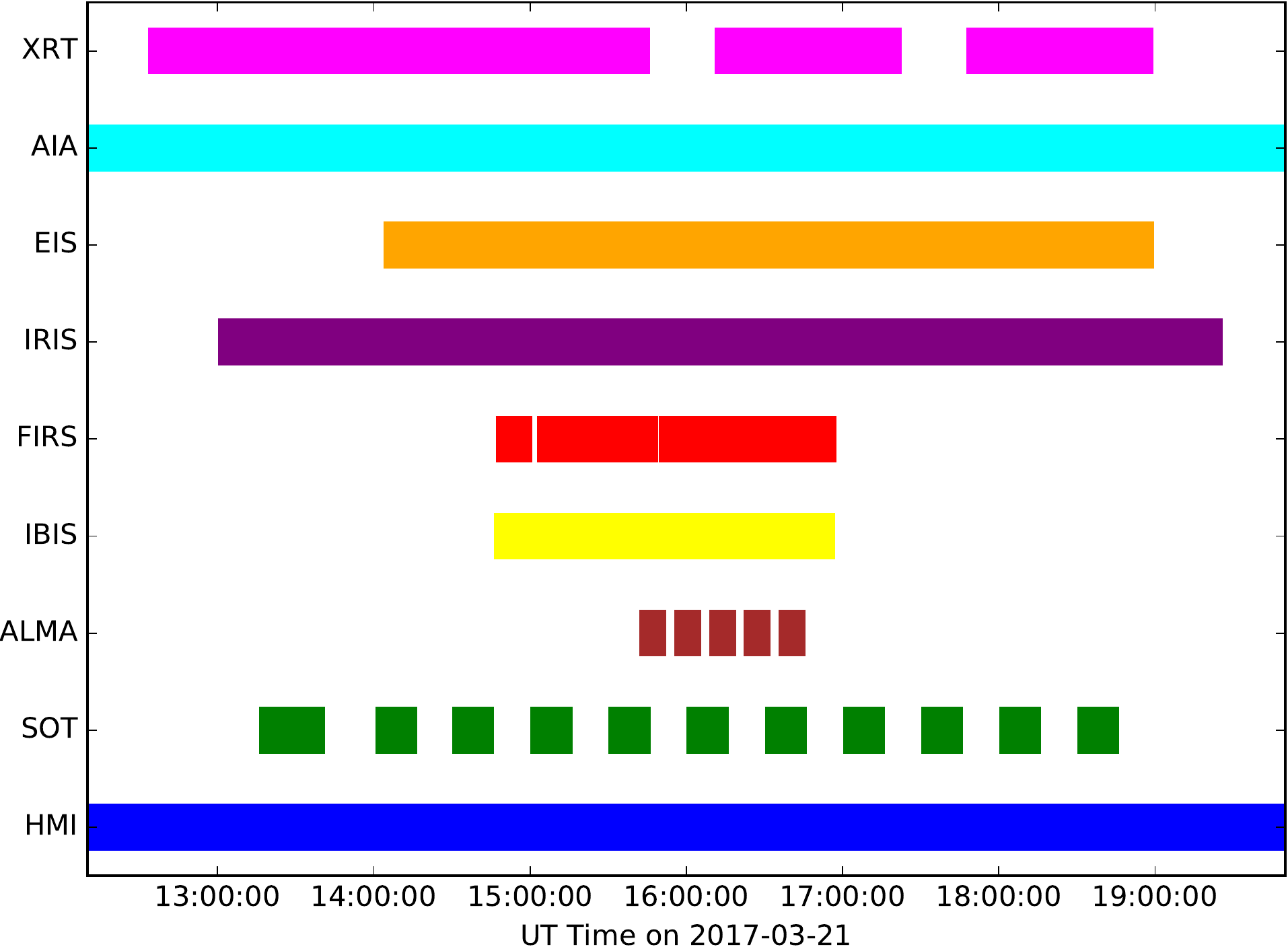}
    \includegraphics[trim={0 17 0 15},clip, width=0.92\linewidth]{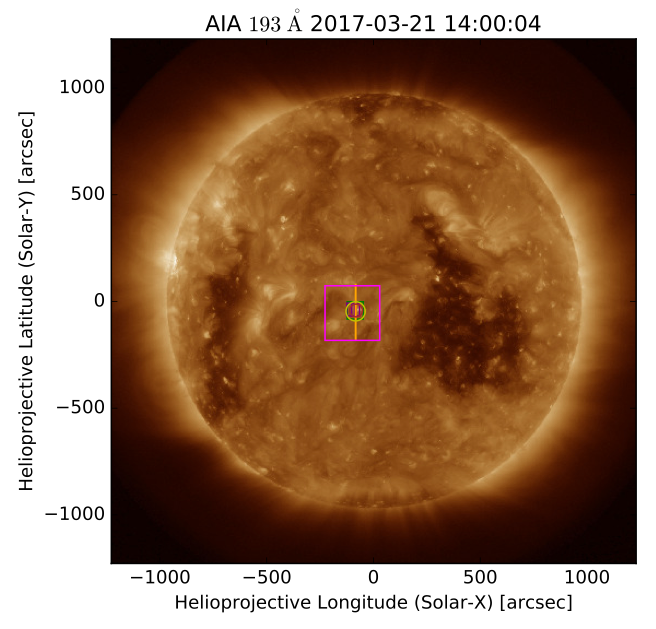}
    \includegraphics[trim={0 0 0 12 },clip,width=0.92\linewidth]{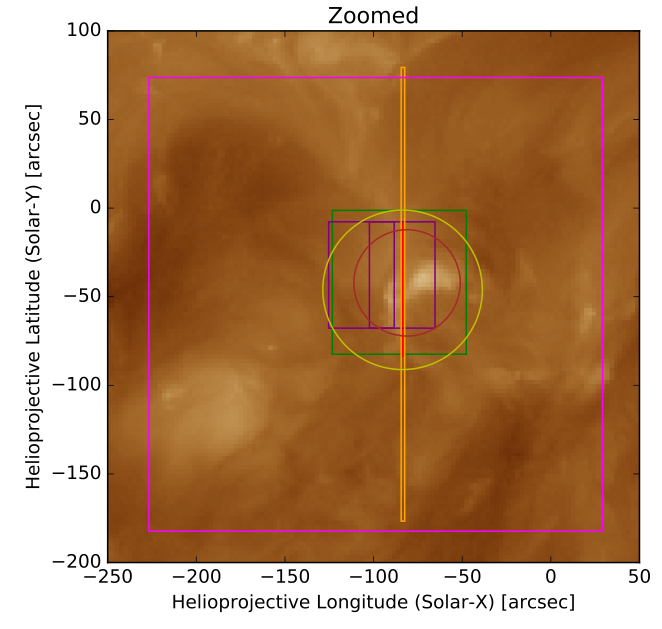}
    \caption{Overview of all observations around the ALMA observing window.  Upper plot: timing information, where instruments are ordered vertically roughly according to height in the solar atmosphere.  Gaps in each color band indicate data gaps in that data channel.  Lower panels: approximate FOVs for each instrument using the same colors as in the upper panel, overlaid on an SDO/AIA 193 \AA{} context image from SOL2017-03-21T
    14:00.}
    \label{fig:coordination}
\end{figure}

The data series presented in this work resulted from a large coordination campaign that was keyed to the ALMA observations and encompassed a suite of ground- and space-based facilities operated by different organizations.
The observations occurred during one of the first coordinated campaigns to support PI-led solar ALMA observations. Due to constraints in the ALMA scheduling, the exact timings and pointings could only be determined the day before the actual observations, with the result that there was not always complete temporal or spatial overlap between each facility.
Given these constraints, our data set represents an enormous success in coordination during this early phase of ALMA solar observations.

Figure \ref{fig:coordination} presents an overview of the temporal and spatial overlap between the various instrument channels during the time of the ALMA observations.
In the upper panel we have ordered the instruments in the vertical direction roughly according to height in the solar atmosphere.
In addition to the times shown here, SOT, EIS, and XRT took observations of this same region for several days leading up to the ALMA observing window, while SDO and AIA provide near-continuous observations in a variety of wavelengths.
The lower panels show the approximate field-of-view (FOV) of each of the instruments overlaid on an AIA 193 \AA{} image using the same colors as the upper panel, with the middle panel showing the full Sun and the bottom panel a version zoomed in on the region.

\begin{figure*}
    \centering
    \includegraphics[width=0.95\linewidth]{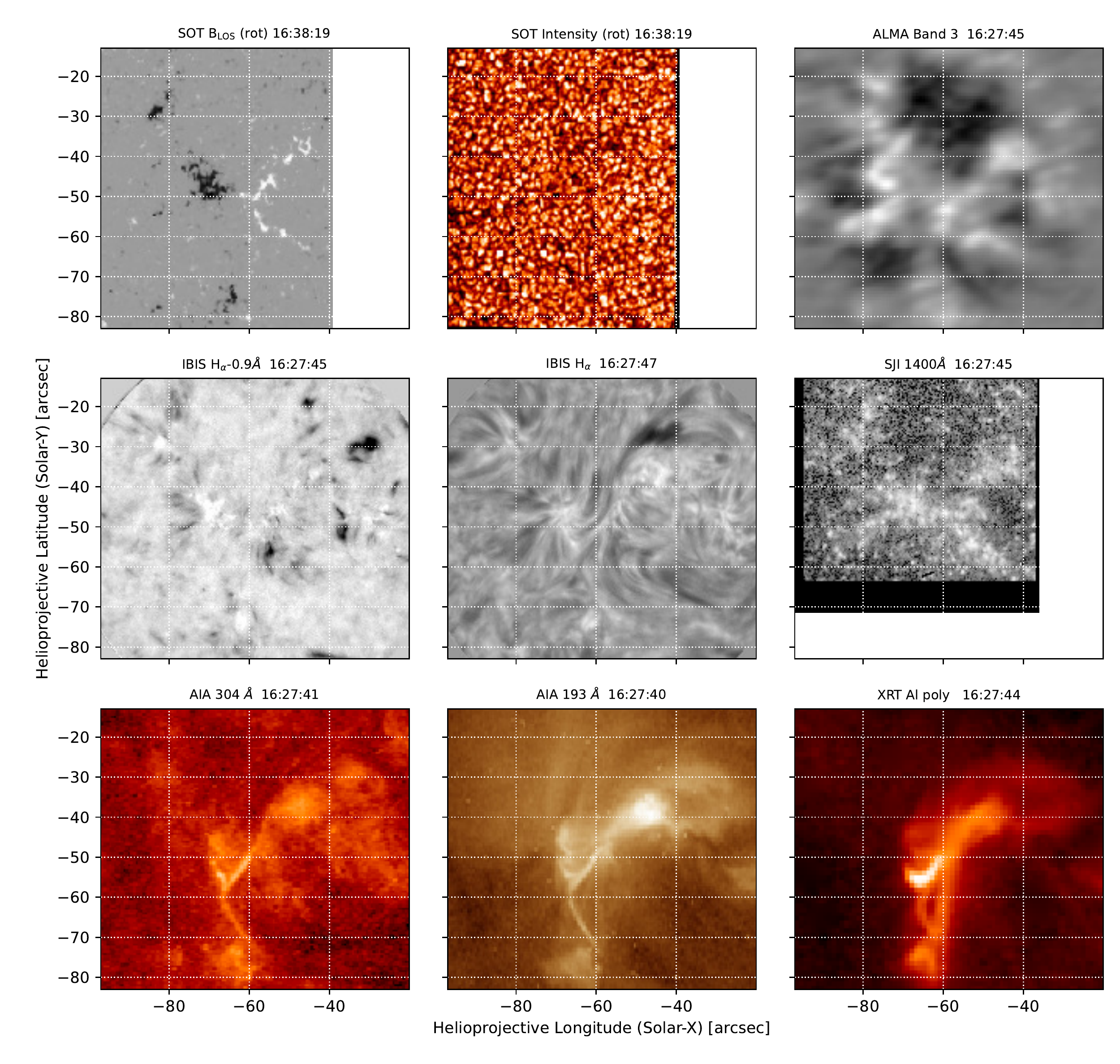}
    \caption{Representative sample of co-aligned images. An animation of this image is available online. The SOT data has been rotated to align with the coordinate time of the IBIS data. The SOT magnetogram (top left) time series shows flux emergence and subduction. As expected, the flux concentrations coincide with weakened granulation patterns in the SOT white light images (top center).  The development and evolution of filamentary structures can be readily seen in the ALMA images (top right). The ALMA data ranges from approximately $\pm$1300 K around {\tt T\_OFFSET}. Short-lived fibril structures appear frequently in the IBIS blue wing (middle left). These fibril structures appear at the ends of the filaments seen in the \halpha{} line center (middle center) and IRIS 1400\AA{} slit-jaw images (middle right). The ends of the features observed in \halpha{} correlate to enhanced brightening in the upper chromosphere and corona, as seen in the AIA 304\AA{} images (bottom left), AIA 193\AA{} (bottom center), and XRT Al\_poly (bottom right). The coronal images show multiple brightenings and the appearance of a thin loop in the southern region of the field-of-view. }
    \label{fig:coordination2}
\end{figure*}

Figure \ref{fig:coordination2} shows a representative sample of co-aligned images from each of the data channels and provides an overview of the target region on the Sun as well as the extent of our coordinated data set.
The images are approximately co-temporal at 16:27 UT and are taken partway though a transient brightening event.
From left to right and top to bottom (and roughly moving from the photosphere through the chromosphere and into the corona) the panels show: 
\begin{itemize}
    {\item (SOT) The photospheric magnetic field including a bipolar region of enhanced network flux}
    {\item (SOT) The solar granulation pattern, including some minor disruptions by the enhanced network concentrations}
    {\item (ALMA) Plasma temperature variations, with hotter/brighter areas roughly coincident with the photospheric magnetic field concentrations}
    {\item (IBIS) \halpha{} blue wing, which shows some dynamic spicules (dark) and the magnetic concentrations (bright)}
    {\item (IBIS) \halpha{} line center, which shows a central sigmoidal filament above the magnetic polarity inversion line}
    {\item (IRIS) Si IV intensity, which shows transient brightenings throughout the FOV, yet mostly concentrated above the enhanced network flux}
    {\item (AIA) 304 \AA{} emission, which shows multiple loop brightenings against background of structured emission}
    {\item (AIA) 193 \AA{} emission showing multiple loop brightenings}
    {\item (XRT) X-ray emission showing multiple transient strands}
\end{itemize}
We discuss this event in more detail in Section~\ref{sec:analysis} below.
NuSTAR also observed this region and detected two small flares around 19:00 and 19:30 UT \citep{2018Kuhar}, confirming continued X-ray activity after the end of our primary coordinated observations, although we do not condsider the NuSTAR observations in the current study.
The remaining subsections describe each data series from the coordinated observations.

\subsection{SDO/AIA}\label{sec:sdoaia}
The Atmospheric Imaging Assembly \citep[AIA,][]{2012LemenEA_AIA} instrument onboard the SDO satellite obtains full-disk images of the Sun every 12 seconds with 0.6\arcsec{} spatial sampling in a variety of visible, UV, and EUV channels.
Observations from all AIA channels are available throughout the entire coordination period.
The Level 1 AIA data was downloaded from the Joint Science Operations Center (JSOC)\footnote{\url{http://jsoc.stanford.edu/}}, updated to Level 1.5 using \fnc{aia\_prep.pro} as described in the SDO Analysis Guide.
We also selected a subregion that fully encompassed the ALMA target using the SolarSoft IDL (SSW) cutout service\footnote{\url{https://www.lmsal.com/get_aia_data/}; see also \S4.3 of the SDO Analysis Guide.} for a region approximately $300\arcsec\times300\arcsec$ for the duration of the other data series (14:00-17:00 UT on 2017-03-21).
The 193 \AA\ channel of AIA was primarily used for co-alignment and to provide full-disk context for the other observations.

\subsection{SDO/HMI}
The Helioseismic and Magnetic Imager \citep{2012SchouEA_HMI} instrument onboard the SDO satellite obtains full-disk images of the Sun in continuum intensity and in polarized measurements of the photospheric Fe I 6173 \AA{} spectral line; the latter is used to determine the doppler velocity, line-of-sight (LOS) magnetic field, and vector magnetic field of the emitting photospheric plasma.  
We used three days of intensity and line-of-sight magnetograms from the level 1.5 \texttt{hmi.M\_720s} and \texttt{hmi.Ic\_720s} data series\footnote{JSOC query: \texttt{hmi.M\_720s[2017.03.19\_00:00:00\_TAI/3d]}, and similar for intensity.}.  
An area $\approx330\arcsec{}\times250\arcsec{}$ was tracked using solar rotation to understand the magnetic evolution of the region leading up to and throughout the coordinated observations.  
To ease later analysis, we derotated all HMI data so that the solar north-south axis aligned with columns of the data array, with north pointing upward, but kept the spatial scaling of the original HMI data.  
The typical extra step of spatially-rescaling the data to match AIA  has little benefit in the present context given that our various data series include many different spatial samplings.

\subsection{ALMA}
\paragraph{ALMA background} The Atacama Large Millimeter Array (ALMA) is a radio interferometer with current frequency coverage spanning 84 to 950 GHz \citep{2009WootenThompson_ALMA}. 
ALMA is constituted of 66 telescope dishes: 54 with a diameter of 12 meters, and 12 with a diameter of 7 meters. 
These 66 dishes are arranged to ensure that ALMA has suitable sensitivity to large-scale diffuse emission, achieved by the high density of centrally located dishes; while also having excellent resolution to small scale features, achieved with the dishes at long distance. 
For this particular observation, ALMA was in its most compact configuration (C43-1), which limits the fine scale resolution, but provides the best available resolution on larger scales. 
The compact configuration also minimizes issues caused by water vapor in the Earth's atmosphere above the telescope.
The ALMA observations presented in this work were in the ALMA observing Band 3, which corresponds to a central frequency of 100 GHz, and a total bandwidth of 18 GHz about that central frequency. 
At this frequency and array configuration, the approximate resolution is 2$\arcsec$ on the sky. 
The integration time was 2 seconds, and the entire observation, including the target field and calibrators, had a duration of 1.03 hours. 

\paragraph{ALMA calibration} The ALMA Band 3 data were cross-calibrated using the standard ALMA pipeline Cycle 4 version\footnote{For a thorough description of the tasks and philosophy see: ALMA Pipeline Team, 2017, ALMA Science Pipeline User’s Guide, ALMA Doc 4.13v2.0.} in CASA version 4.7.2 \citep{casa}. 
The observations were flux calibrated with the strong millimeter source, J2253+1608, approximately 6.5 Jy in ALMA Band 3, used for the bandpass and flux calibration, and the strong millimeter source, QSO\_B0003-066, approximately 2 Jy in ALMA Band 3, used for the phase cross-calibration. 
The initial cross-calibration eliminated 4 dishes due to unreasonably elevated system temperatures, and produced images of a reasonable quality, but still suffered from some poor calibration artifacts.
More details on the specifics of ALMA solar calibration are described in \citet{2017ShimojoEA_ALMASolarInt} and \citet{2017WhiteEA_ALMASD}.

\paragraph{ALMA self-calibration} In order to improve the fidelity of our images we then self-calibrated the data \citep{1984PearsonReadhead}. 
We iteratively applied phase-only self-calibration using the \fnc{CLEAN} algorithm \citep{1974Hogbom} over the entire ALMA data series (including both 12 m and 7 m baselines) until there was minimal out-of-field r.m.s. noise and the image showed no improvement upon further iteration.
This process minimized the PSF pattern relics in the image and sharpened the image, which is why this process if often compared to the effect of focusing a telescope or microscope.\par

For unknown reasons, the phase-center of these specific observations was shifted from the expected beam center, causing an apparent misalignment between the center of the field of view of and the contrast center of the observation.
Due to the co-alignment methods described below, this is not of particular concern, but is relevant in case of re-calibration.\par

We provide the final calibrated data in FITS format stored as a flux (Jy/beam), but the linear conversion to Kelvin has already been calculated and stored in the FITS headers under the keyword \texttt{FLUX2T}.
The zero-point brightness temperature \texttt{T\_OFFSET} is set to the value of 7300 K determined for disk center quiet Sun by \citet{2017WhiteEA_ALMASD}, which should well represent this data series. This offset is not applied in Figures~\ref{fig:coordination2},~\ref{fig:subset},~and~\ref{fig:wishbonezoom}.

\subsection{DST/IBIS}
The Interferometric BIdimensional Spectrometer \citep{2006Cavallini_IBIS} is a dual Fabry-Perot interferometer- (FPI-) based imaging spectrometer that was mounted at the Dunn Solar Telescope.
The FPI cavity is tuned to transmit a narrow bandpass $(\approx 2.4\unit{pm})$, and the tuning is modulated to step the bandpass over the wavelength range of the spectral line, in this case the hydrogen alpha line (\halpha{}) at $656.3\unit{nm}$.
We used IBIS in spectral-only mode (no polarimetry) with a circular field stop of approximately 90\arcsec{} diameter.
We continuously sampled the \halpha{} line between 14:46-16:55 UT using 26 wavelength positions, with $\approx 12.5\unit{pm}$ spacing near line core and $\approx 19.1\unit{pm}$ spacing in the wings.  
Each narrow band image is paired with a strictly co-temporal broad band continuum image taken in a nearby wavelength band at $660\unit{nm}$.
The broad band image is used for speckle reconstruction, self-alignment of the narrow band images, and co-alignment between the IBIS data and other data series, e.g., to the HMI continua data series. \par

The IBIS data were reduced primarily using the pipeline code provided by NSO\footnote{Version 1.4, available here: \url{https://www.nso.edu/telescopes/dunn-solar-telescope/dst-pipelines/}.}.  
This process aligns the broad band and narrow band channels and accounts for detector dark counts, flat field, and the spatially dependent wavelength shift induced by the telecentric mount of the FPIs.  
Kevin Reardon provided several modified steps that better account for spatially and spectrally dependent fringes in the narrow band data due to the prefilter, as compared with the pipeline code\footnote{Reardon, private communication.  See documentation at \url{https://github.com/kreardon/IBIS.git}.}, which resulted in an improved estimate of the gain for each spatial and spectral point throughout the narrow band datacube.  
As a final calibration step, we used the KISIP code \citep{2008WoegerSPIE,2008WoegerEA_Speckle} to perform a speckle reconstruction of the data using the IBIS broad band images that are taken strictly co-temporally with the narrow band images; the broad band data is specifically designed for this purpose \citep{2008Cauzzi}.\par

We took several further steps for the IBIS preparation.  
To correct for the several arcsecond inaccuracy of the DST blind-pointing, slight rotation of solar north with respect to the CCD pixel arrays, and the slight difference in plate scale between the CCD X and Y directions, we rotated, shifted, and stretched the IBIS broad band data to match the granulation pattern in nearly co-temporal HMI continuum data at time 2017-03-21 15:46:36 UT.
These parameters remained constant throughout the IBIS data series.
\par

\subsubsection{IBIS self-alignment}\label{sec:ibisselfalign}
IBIS data in the spectral dimension is not strictly co-temporal: the time difference between successive wavelength steps is about $0.167\unit{s}$ and the cadence between images at a given wavelength is about $4.2\unit{s}$.
Our observations were taken during periods of moderate seeing, so the image sequence has some jitter, which became significant at times.  
To correct for this effect and spatially co-align all the IBIS data with itself we used a cross-correlation technique outlined here: 
\begin{enumerate}
    {\item We manually determined ``bad frames'' when the AO lock was lost or a frame became significantly distorted, by inspecting the strictly co-temporal broad band images; these frames were assumed bad for all wavelengths in a given scan of the line.}
    {\item We generated a reference image by taking the running average of the previous 5 registered ``good'' scans through the spectral line: at 26 images per scan, the average involves 130 broad band images.
    The initial average reference image was taken to be the average of the first spectral scan from the data series.}
    {\item We used \texttt{chi2\_shift}\footnote{The \texttt{chi2\_shift} method from Adam Ginsburg's \texttt{image\_registration} python package.  See \url{https://github.com/keflavich/image_registration}.} to determine the spatial offset between each of the broad band images in a scan and the average reference image.}
    {\item If the frame was labeled ``good,'' the final registered broad band image for each wavelength was appended to running average.}
    {\item The measured offset was then saved, to be applied later to both the broad band image and the co-temporal narrow band image, which will co-align the entire data series (this last step has already been performed on the provided IBIS FITS files).}
\end{enumerate}

The above method allowed essentially all 46,800 frames to be co-aligned while being robust against the occasional loss of AO during periods of poor seeing.\par

\subsection{DST/FIRS}
The Facility InfraRed Spectropolarimeter \citep{2010JaeggliEA_FIRS} is a high dispersion dual-beam spectropolarimeter with the ability to operate simultaneously at visible (6302 \AA) and infrared (10830 or 15650 \AA) wavelengths.  
FIRS has a scanning mirror and a variety of reflective slit units that can provide optional multi-slit capability for highly efficient raster scans of the solar surface.  
The light reflected from the mirrored slit unit can be reimaged to provide context during observations.  
As a facility instrument at the Dunn Solar Telescope, FIRS can receive a seeing stabilized image provided by the high order adaptive optics system.

During the coordinated observations, FIRS first conducted a raster observation from 14:46:47 to 15:00:54 UT (08:46:47 to 11:00:54 am MDT).  Then sit-and-stare observations were run almost continuously from 15:02:36 to 16:57:42 UT with small gaps for adjustment of the adaptive optics system.  During these observations, FIRS was configured to use the 40 $\mu$m single slit with the f/36 feed optics, which provide a slit width of 0.3 arcsec on the sky.  The vertical extent of the slit covered approximately 74 arcsec.  Only the 10830 \AA\ channel of FIRS was used, and the narrow band filter was removed to take advantage of the maximum wavelength coverage possible with the detector.  In this configuration the spectrograph was able to cover a 40 \AA\ bandpass centered at 10834 \AA.  This region contains the Si I and triplet He I lines commonly used for photospheric and chromospheric  spectropolarimetry, as well as several other solar and telluric lines.  The spectral sampling of 3.86 pm/pixel is approximately equal to the spectral resolution of the instrument at this wavelength, based on laser profile measurements \citep{2011Jaeggli_PHD}.  An exposure time of 125 msec was used to keep counts within the range of linear behavior for the HgCdTe detector.  The liquid crystal variable retarders ran two repeats of a 4-state modulation sequence, and the time to execute a complete polarization measurement was about 4 seconds.

A camera was set up to reimage the light from the mirrored surface of the FIRS slit.  These slit-jaw images were obtained to assist in co-alignment and cover a $185\times155$ arcsec$^2$ field of view with 0.153 arcsec/pixel sampling.  The slit-jaw images were recorded from 14:46 to 16:58 UT with a cadence of 5 seconds to approximately match the cadence of the FIRS spectrograph, although the exposure time was only 200 msec.  The wavelengths seen by the slit-jaw imager were longer than about 700 nm due to the beam splitter needed for the IBIS \halpha{} channel, and the sensitivity of the silicon-based detector for the slit-jaw imager extends out to about 900 nm.

The data from FIRS were reduced using techniques similar to \citet{2012Jaegglietal}.  First, calibration data were assembled.  All images were first corrected for the non-linear response of the detector.  All frames from a single dark calibration were averaged together, and the nearest dark calibration in time was applied to the frames with light.  A raster scan of the grid target at the telescope main focus was used to determine the geometric correction to make the spectral and spatial coordinates orthogonal with linear dispersion, and to match the coordinates between the dual beams for spectropolarimetry.

A flat field for the science data was constructed using an observation of disk center with randomized motion to blur out spatial features.  The geometric correction was used to convert the spectrum to spatial/spectral coordinates from detector coordinates.  The average spatial and spectral profiles were obtained.  The hairlines crossing the slit were fit from the spatial profile and divided from the image so that these features would remain in the science observations.  The spectral lines were fit from the spectral profile using a Voigt function and then divided from the flat field observation.  The spectral profile was then transferred back to detector coordinates and divided from the original to produce a master solar flat without spectral lines.

A pixel-by-pixel polarimetric correction using the method of \cite{2013Schad} was attempted using a calibration sequence obtained with the ASP calibration linear polarizer and waveplate near the DST main focus.  However, the derived solution did not seem to properly correct the science data, leaving large bias levels in the polarized states.  This may be due to changes in the IR detector linearity that were not properly corrected for by the lookup table.  Instead, a blind polarimetric demodulation was applied to the data, not correcting for instrumental polarization.  The resulting polarized spectra contain mostly Stokes V signal.  Application of ad-hoc techniques for determining and removing instrumental polarization, i.e. \citet{collados03}, would probably not be successful due to the lack of strong linear polarization signatures.  Magnetogram-style maps of the net polarized signal in the Si I line appear similar to the SOT/SP maps.

Fiducials crossing the slit provide common features for co-aligning the slit in the vertical and horizontal directions with respect to the image obtained by the slit-jaw camera.  During the reduction for the spectrograph data, the dark fiducial lines crossing the spectra were fit in each spectrum and then divided to remove them from the image.  The slit and fiducials were fit in each slit-jaw image before application of the flat field.  The fitted fiducial and slit positions were saved for the co-alignment step.

\subsection{IRIS}
The Interface Region Imaging Spectrograph \citep{2014DePontieuEA_IRIS} is a UV spectrograph with a slit-jaw imager that provides co-temporal imaging and spectra in several UV bandpasses.  
During the period 13:00:07 to 19:26:15 UT on 2017-03-21, IRIS performed 530 repeats of a program taking coarse 8-step rasters using a medium field of view with the slit aligned north-south.  
The spectrograph field of view, approximately 60 $\times$ 16 arcsec$^2$, was sampled in 2 arcsec steps by the 0.3 arcsec wide slit, while the silt-jaw imager had a 60 $\times$ 65 arcsec$^2$ instantaneous field of view centered on the spectrograph slit.  
The spectrograph took 4 sec exposures at a 5.4 sec cadence with both the FUV and NUV spectrograph channels. 
The slit-jaw imager obtained images every 11 sec using the Si IV 1400 \AA\ filter channel, and images with the 2832 \AA\ Mg II line wing filter channel were taken every 44 sec to provide photospheric context.  
IRIS observations were continuous in the time period and include several passages through the South Atlantic Anomaly, during which the images show increased hot pixels due to particle strikes.

\begin{figure*}
    \centering
    \includegraphics[width=7in]{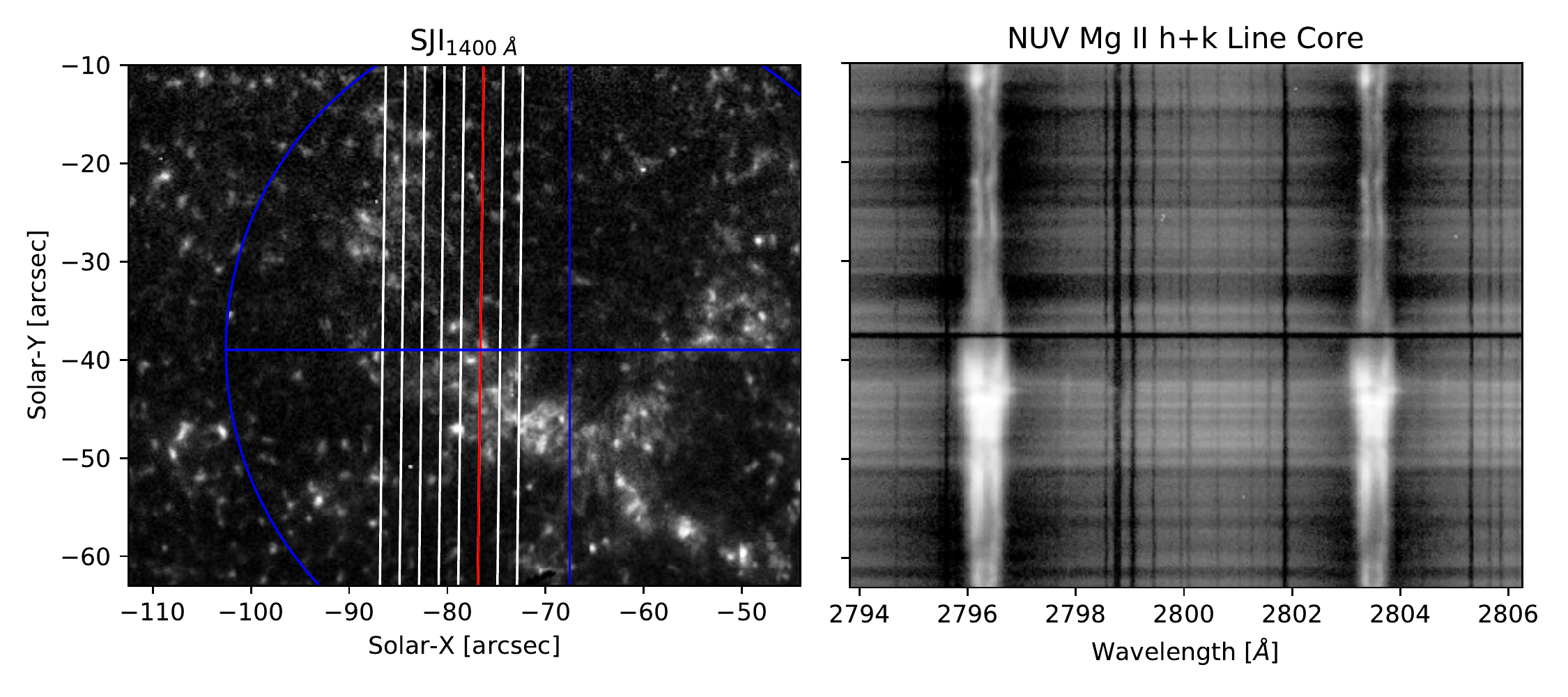}
    \caption{An example of the IRIS Level 2 SJI and NUV spectrograph data near the time 15:47 UT.  The slit-jaw image is a composite of 4 images taken from 15:46:57.57 to 15:47:30.35 UT on 2017-03-21 to show the full field of view during the raster.  The slit positions for the 8-step coarse raster are shown over plotted on the SJI image in white.  The red slit indicates the position at which the NUV spectrum on the right was taken at 15:47:25.03 UT.  The blue circle and crosshair shows the co-rotating ALMA field of view.}
    \label{fig:IRIS}
\end{figure*}

IRIS Level 2 data was obtained from the IRIS website.  
Calibration of the IRIS data are described in \citet{wuelser18}.  
IRIS Level 2 data is already co-aligned with AIA to a high degree of accuracy, so further co-alignment steps were not necessary.  
An example of the IRIS SJI Si IV 1400 \AA\ image and NUV spectrum near the line core are shown in Figure \ref{fig:IRIS}.

\subsection{{\it Hinode} SOT}
The Solar Optical Telescope Spectropolarimeter (SOT/SP) is a slit spectrograph with polarimetric capabilites onboard the {\it Hinode} satellite.
It measures the full Stokes polarization of the Fe I lines at 630.15 and 630.25 nm, which form at photospheric heights and are sensitive to the Zeeman effect.
SOT/SP data was primarily taken from 14:00 UT to 18:46 UT with using 60\arcsec{} $\times$ 60\arcsec{} fast maps with 0.3\arcsec{} spatial resolution with a $\sim$ 30 minute cadence. 
Additionally, the region of interest was observed sporadically over the 50 hours leading up to the coordinated observations, with FOVs of approximately $150\arcsec{}\times160\arcsec{}$.  
The SP instrument performance is described in \citet{2013Litesetal}.

The SP Level 2 data, reduced and inverted with the Milne-Eddington gRid Linear Inversion Network (MERLIN) code to produce maps of physical atmospheric parameters, including the vector magnetic field, were obtained from the Community Spectropolarimetric Analysis Center (CSAC, DOI:10.5065/D6JH3J8D).  See \citet{2013Lites&Ichimoto} for a description of the data reduction routines.

\subsection{{\it Hinode} EIS}
\begin{figure}
    \centering
    \includegraphics[width=3.5in]{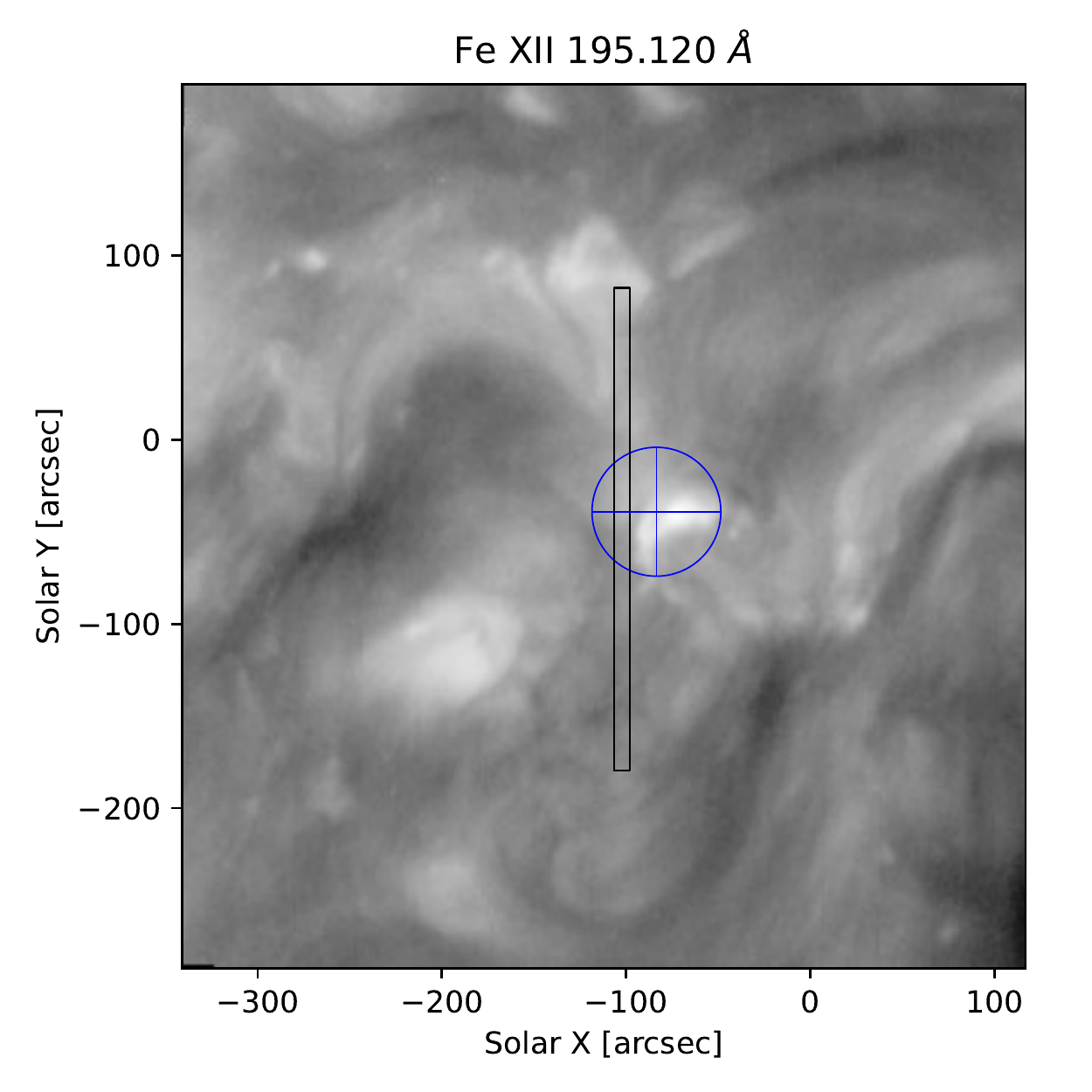}
    \caption{A reconstructed slot raster image from the EIS \ion{Fe}{12} 195 \AA\ channel. The slot raster started at 14:00 UT.  The blue circle and crosshair show the ALMA co-rotating field of view.  The black rectangle shows the extent of the region covered by the EIS slit during the sit and stare observations.}
    \label{fig:EIS}
\end{figure}

The EUV Imaging Spectrometer \citep{2007CulhaneEA_EIS} onboard the {\it Hinode} satellite provides slit spectra covering many spectral lines in the extreme ultraviolet range, between 170 and 290 \AA{}.  
EIS conducted observations for the coordinated campaign from 14:00 to 21:21 UT on 2017-03-21. 
This instrument provides a wide range of diagnostics as described in \citet{2007Young}. 
Context slot rasters approximately three minutes in duration were taken at the beginning and end of the observation.  
An example slot raster is shown in Figure~\ref{fig:EIS}.
In between the slot rasters, EIS took sit-and-stare observations with the 2\arcsec{} slit at a cadence of 34 s. 
The sit-and-stare observations were interrupted by a 10 minute disk center synoptic observation at 19:40 UT.

The EIS slot rasters covered a large region approximately $470\arcsec\times485\arcsec$ in 15 overlapping steps with spatial sampling of 1\arcsec\ per spatial pixel.  
The sit-and-stare observations consisted of a 20\,s exposures with the 2\arcsec\ wide slit with a spatial sampling of 1\arcsec\ per pixel along the slit. 
The detector readout for the spectra included all of the available wavelength ranges, but was limited to 256\arcsec\ along the slit. 
Level 0 EIS data were downloaded via the Virtual Solar Observatory\footnote{https://sdac.virtualsolar.org/cgi/search} and processed using the default settings for the \verb+eis_prep+ routine in SSW \citep{1998FreelandHandy_SSW}.  
Intensities for the \ion{Fe}{12} 195.119\,\AA\ and \ion{He}{2} 256.317\,\AA\ lines were computed using multi-Gaussian fits with the MPFIT IDL package \citep{2009Markwardt}.

\subsection{{\it Hinode} XRT}
The X-Ray Telescope \citep{2007GolubEA_XRT} onboard the {\it Hinode} satellite uses grazing incidence optics to obtain thermal X-Ray images of the solar corona.
XRT observed the active area from 12:33:12.311 -  18:59:27.511 UT with the Al\_poly filter, enabling plasma observations above $\sim$3 MK. The Al\_poly filter is one of the thinnest filters on XRT which has not been prohibitively impaired by contamination \citep[][]{2011NarukageEA_XRT3}, though some features of the contamination are still apparent. 
While the intended cadence was 4\,s, automatic exposure control increased the exposure time, which slowed the cadence to 16\,s per image. 
This large data series was prepped \citep[][]{2014KobelskiEA_xrtprep} and the methods of \citet{2015YoshimuraMcKenzie_XRTcoalign} were used to estimate offsets for co-alignment to AIA (as illustrated in \figref{fig:coalign}).
Cross-correlation techniques \citep[such as {\tt tr\_get\_disp.pro} in SSW;][]{1998FreelandHandy_SSW} were utilized to further remove instrumental jitter and to self-align the XRT data with itself.

\section{Co-alignment} \label{sec:coalign}
\begin{figure}[b]
    \centering
    \includegraphics[width=1.0\linewidth]{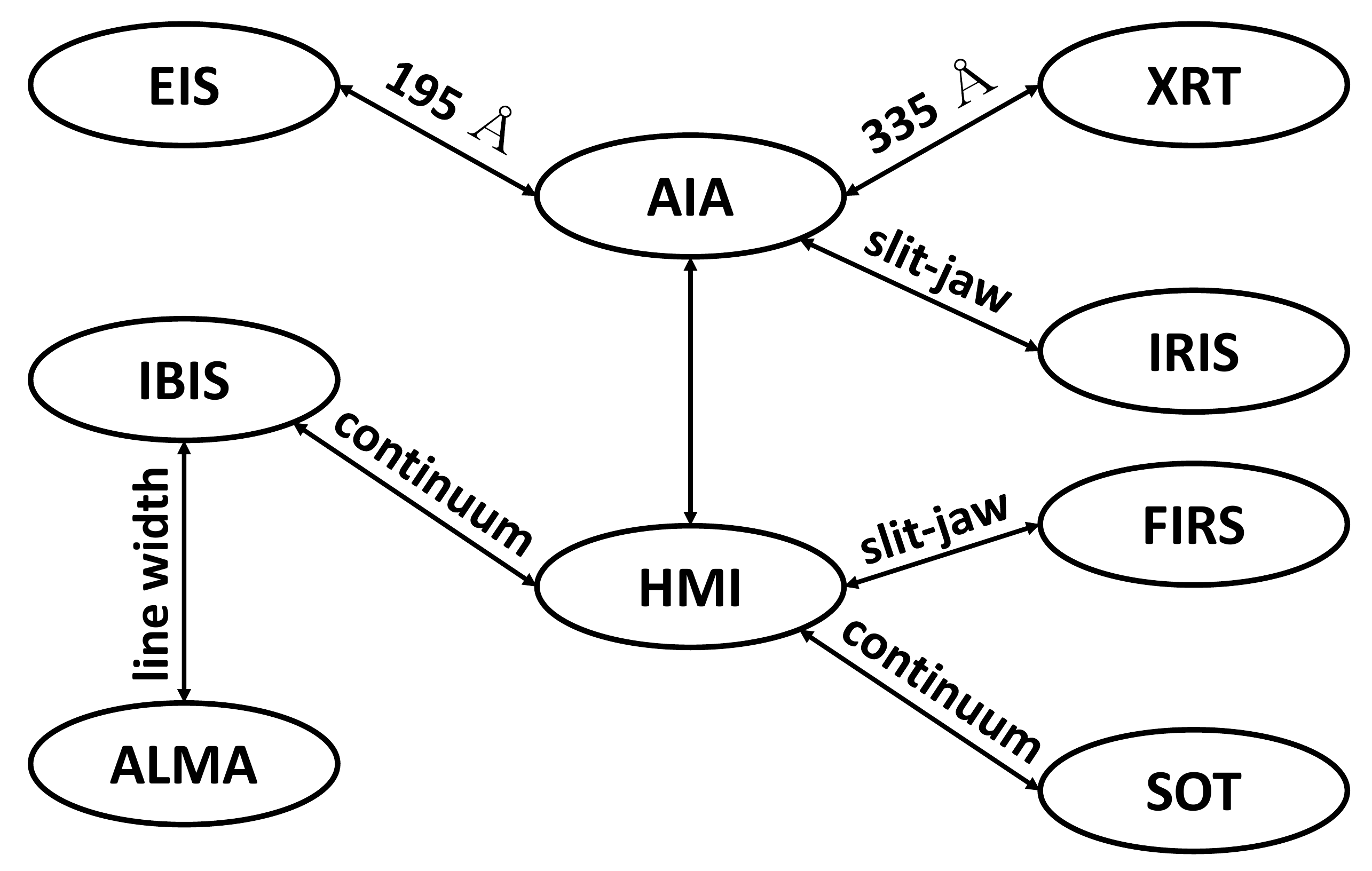}
    \caption{Graphical representation of the co-alignment process. SDO was utilized as the central alignment reference, with other instruments co-aligned to it as possible. }\label{fig:coalign}
\end{figure}

Co-aligning the various instruments was a challenge.  
\figref{fig:coalign} graphically depicts how each data series was co-aligned.
Our general approach was to use the broad wavelength and height coverage of the various HMI and AIA channels as a ``ground truth'' mapping between solar coordinates and morphological features seen within each data series.
The 2012 Venus transit allowed subarcsecond alignment of all the AIA and HMI channels\footnote{See the SDO Data Analysis Guide at \url{https://www.lmsal.com/sdodocs/doc/dcur/SDOD0060.zip/zip/entry/}, \S{}7.1 in the Sept 14, 2020 version of the document.}, which has since been maintained using Mercury transit observations, making SDO an excellent resource for this task.  
Alignment was then verified by checking against data series that had not been explicitly aligned with each other.
For example, the IBIS broadband images were aligned to HMI continuum data and then that alignment was verified by checking the correspondence of enhanced emission in the \halpha{} line wings to the HMI magnetograms, SOT magnetograms, and UV emission in the 1600 and 1700 \AA{} AIA channels. 

Each data series has different temporal and spatial samplings and physical extent, so no attempt was made to resample images from each instrument to a common temporal/spatial grid.  
Instead, the reduced and co-aligned data series make use of the World Coordinate System (WCS) variant for solar physics defined in \citet{2006Thompson} and included in the FITS headers of our public data set.  
All data in this work uses helioprojective-cartesian coordinates, which can be transformed to any other coordinate system using standard routines in \fnc{SSW}, \fnc{astropy}, or \fnc{sunpy}.\par

\subsection{DST/FIRS}

During the reduction of the FIRS slit-jaw images, the FIRS slit and hairlines crossing the slit were fit in each image before application of the flat field.  The pixel coordinates of the slit/hairline pairs were saved for use during the co-alignment step.  The dark hairlines also appear in the FIRS spectrograph images.  During the reduction of the spectrograph data, the positions of these were also determined and saved for later use.

The reduced FIRS slit-jaw images were co-aligned to SDO/HMI intensity images from the 45\,s series.  The approximate center coordinates of the FIRS slit-jaw were taken from the image headers which contain the telescope pointing near the time of the observation.  The spatial dispersion was estimated based on observations of the grid target.  The nearest HMI image in time was taken, and the coordinates were rotated to the time of the FIRS slit-jaw observation.  Using interpolation based on the estimated slit-jaw coordinates, a sub-field was extracted from the HMI image.  X and Y shifts between the HMI sub-field and the slit-jaw image were determined using the SSW routine \verb+tr_get_disp+.  The spatial dispersion and rotation of the first image was adjusted by hand to achieve a good match, then the co-alignment of the entire image sequence was done using the same scale and rotation parameters.  The resulting coordinate mapping was used to transform the measured slit and hairline crossings into solar coordinates.  The FIRS slit-jaw images were written to FITS files with their updated coordinate information using the WCS standard.

To get the FIRS spectrograph data into the solar coordinate frame, the coordinates of the slit/hairline crossings for the slit-jaw data series were interpolated at the time of FIRS spectrograph observation.  The slit/hairline solar coordinates were then used to determine the spatial dispersion along the slit, the solar X and Y coordinates at the center of the slit, and the rotation of the slit with respect to solar north assuming a linear mapping.  Each set of polarized spectra was then written to a FITS file with the spatial and spectral coordinates in the WCS standard.  The Stokes I spectrum was written in the main extension, and the Stokes Q, U, and V states were written to subsequent extensions with abbreviated headers.

\begin{figure*}
    \centering
    \includegraphics[width=5in]{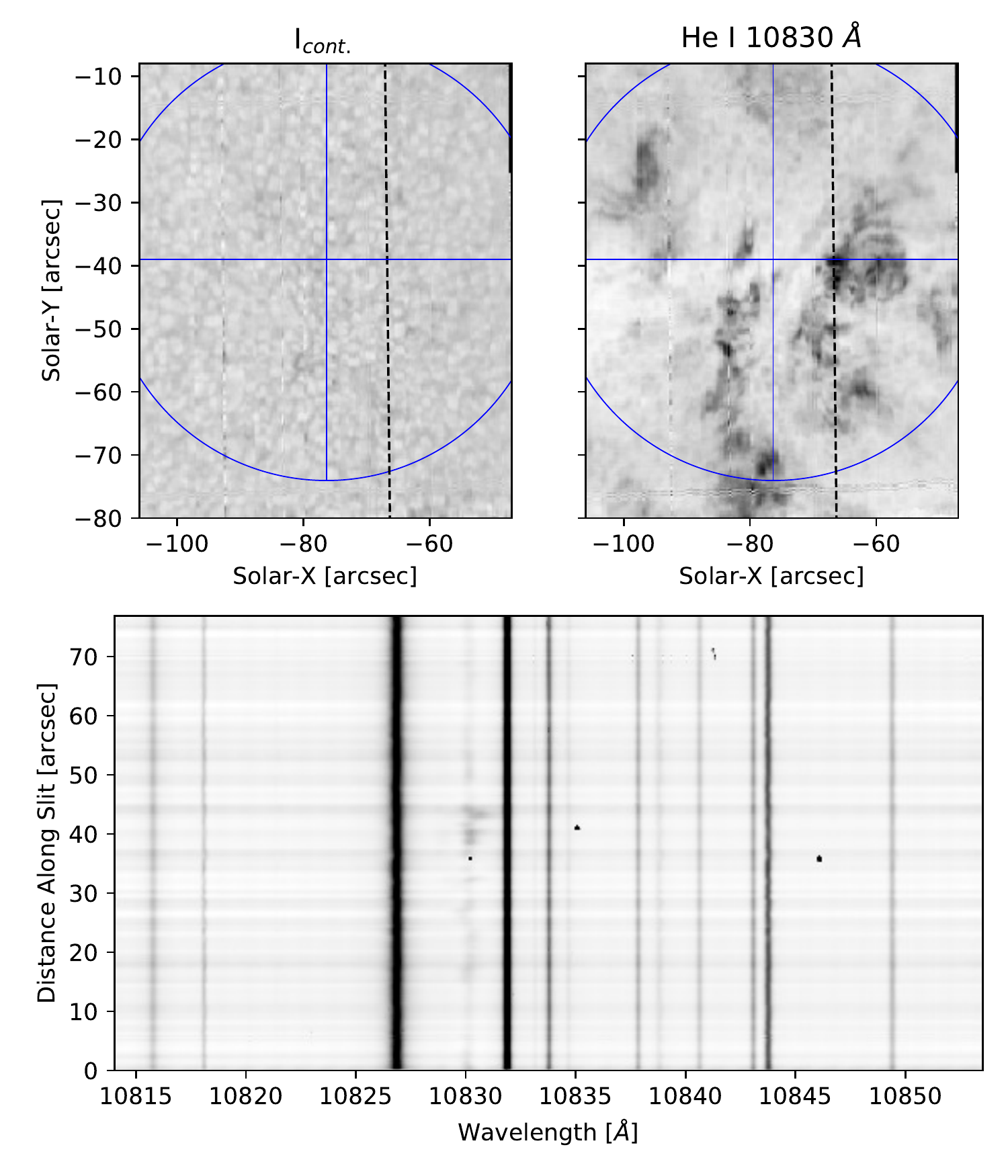}
    \caption{co-aligned FIRS spectograph data from the beginning of the coordination taken between 14:46:48 and 15:00:55 UT on 2017-03-21.  The top two panels show the reconstructed maps of continuum and He I 10830 line intensity (left and right respectively).  The bottom panel shows the full Stokes I spectrum obtained with FIRS at 14:53:53 UT at the position indicated by the dashed line in each of the maps.  The blue circle and crosshair shows the co-rotating ALMA field of view.}
    \label{fig:FIRS}
\end{figure*}

Representative images and spectra can be found in Figure~\ref{fig:FIRS}.

\subsection{{\it Hinode} SOT}

\begin{figure*}
    \begin{center}
        \includegraphics[width=7in]{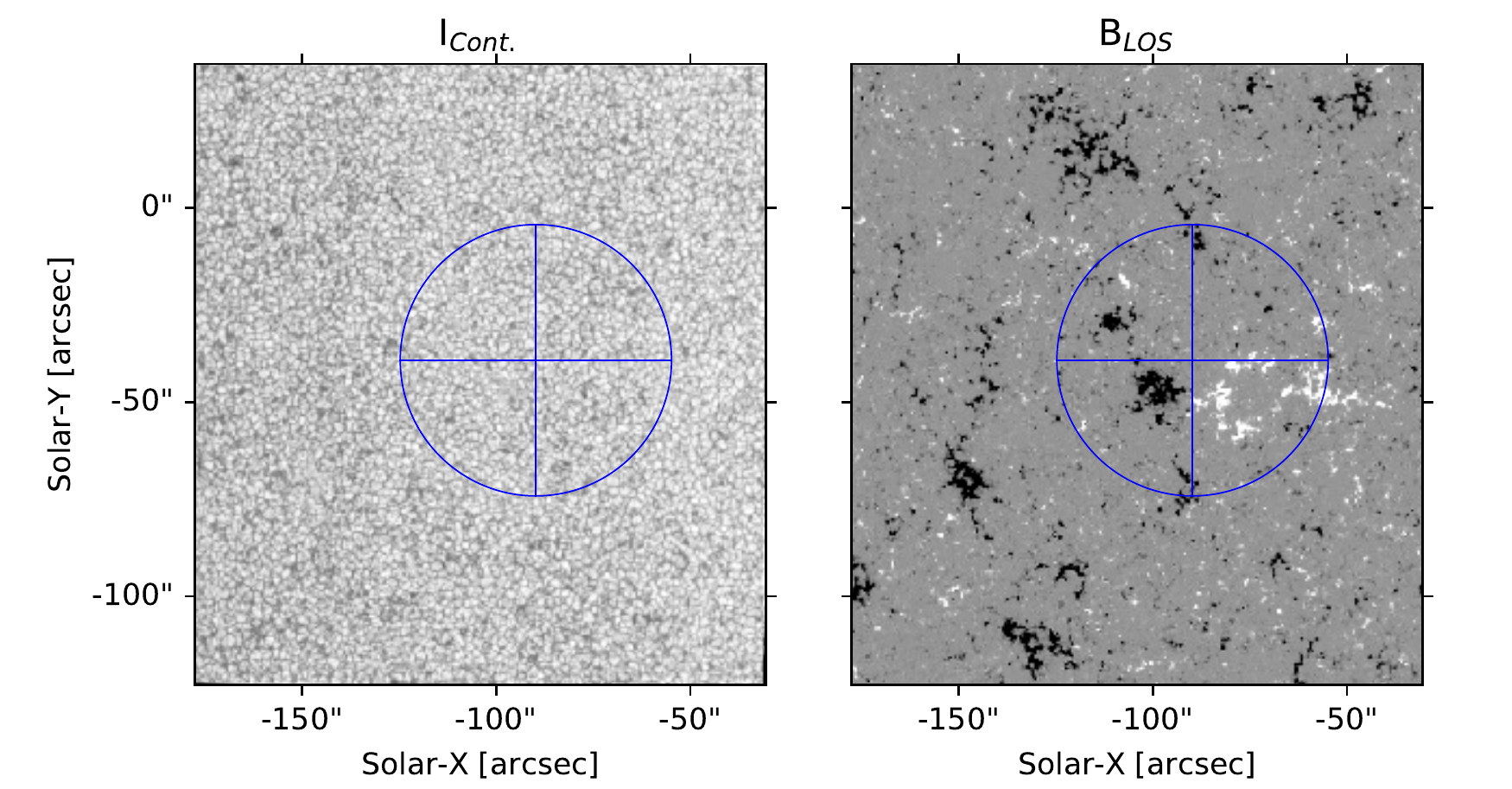}
    \end{center}
    \caption{The co-aligned continuum intensity and line of sight magnetic field from the co-aligned Level 2 SOT/SP data based on inversion of the 630.2 nm Fe I lines.  The raster scan was run from 13:16:01 to 13:48:21 UT on 2017-03-21 and was the last large raster taken before the coordinated sequence began.  The blue circle and crosshair shows the co-rotating ALMA field of view.}
    \label{fig:SOTSP}
\end{figure*}

For each Level 2 raster scan of SOT/SP, the solar coordinates of the slit at each raster position were differentially rotated from the observed time to a common time at the center of the scan (using the SSW routine \verb+drot_xy+).  The parameter maps were then resampled to a uniform coordinate grid with the same spatial sampling as the original data.  HMI intensity observations closest in time to the center of the SOT/SP scan were selected and interpolated onto the same coordinate grid, and the relative shift between HMI and the SOT/SP continuum intensity map was determined (using \verb+tr_get_disp+).  The SOT/SP coordinates were updated with the corresponding shift.  An ad-hoc rotation angle of CROTA2=-0.6 deg and x-y pixel scale of 0.315 arcsec/pixel were determined with respect to HMI and applied simultaneously with the coordinate shift.  The resampled and co-aligned SOT/SP parameter maps were written to WCS-compliant FITS files where the main extension contains continuum intensity, and the inverted parameters are contained in additional FITS image extensions.  Figure \ref{fig:SOTSP} shows an example of the continuum intensity and line of sight magnetic field for the last large raster taken before the coordination with ALMA commenced.

\subsection{{\it Hinode} EIS}
\begin{figure}
    \centering
    \includegraphics[width=0.85\linewidth]{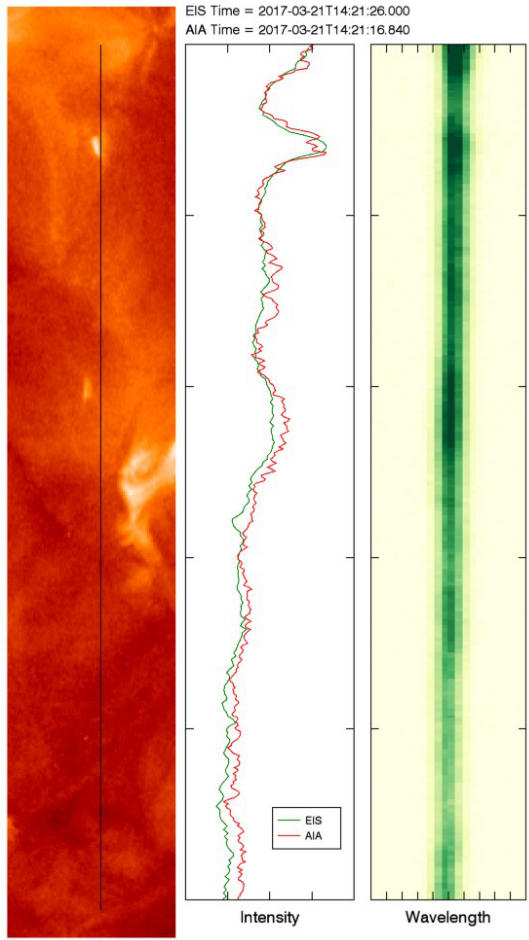}
    \caption{Co-aligned EIS and AIA data. The left panel shows an AIA 193\,\AA\ image with the position of the EIS slit indicated by the black line. The AIA image was taken at 14:21:16.840 UT, and the EIS image from 14:21:28.000 UT. The center panel shows the EIS and AIA intensities along the slit. Note that the EIS intensities here are computed by convolving the spectrum with the AIA effective area. Intensities derived from Gaussian fits to the line profile were also computed. The right panel shows the EIS exposure in the vicinity of the \ion{Fe}{12} 195.119\,\AA\ line. }
    \label{fig:eis_aia}
\end{figure}

The EIS slit data were co-aligned to 12\,s, $202\arcsec\times456\arcsec$ 193\,\AA\ AIA cutouts obtained from JSOC. For each EIS exposure we convolved the full CCD spectrum with the AIA effective area for the 193\,\AA\ channel. We also resampled the AIA images to better match the cadence and plate scale of the EIS data. We then cross correlated the EIS intensities along the slit with the the intensities in the resampled AIA image taken closest in time to the EIS exposure. We computed correlations over a range of positions close to the commanded pointing and recorded the position with the highest correlation. An example of co-aligned EIS and AIA intensities is shown in Figure \figref{fig:eis_aia}.

The EIS slot rasters at the beginning and end of the observations were also co-aligned to AIA 193 \AA\ images, but using a slightly different technique.  The EIS 195 \AA\ window was extracted for each slot position in the raster and the nearest AIA 193 \AA\ image in time was taken.  The EIS metadata values for rotation and spatial sampling of the EIS slot were adjusted, to $1^\circ$ and $1.00\arcsec{} \times 0.99\arcsec{}$ respectively, for all slot data, and the center field position of each slot pointing was determined by eye to achieve a good match.  This rough pointing information was used to generate 2D helioprojective Cartesian coordinate arrays for the x and y coordinates of the image.  The AIA data was interpolated to the EIS coordinates, then the SSW routine \verb+tr_get_disp+ was used to determine the residual shift between the EIS and AIA images in an automated way.  These pixel shifts were then used to update the EIS coordinate arrays.

FITS files containing the EIS slot raster and spectrograph sit-and-stare co-aligned coordinates arrays are available with the data release accompanying this publication.  
The coordinate arrays do not strictly comply with the WCS standards used for the other data series because of the irregular format of the Level 1 EIS data set.  
Altering this data set from its original format would invalidate the wealth of routines that are available for EIS data analysis.  
The co-aligned coordinates are only valid for the EIS 195 \AA\ wavelength because EIS has a systematic vertical shift with wavelength due to a small tilt of the grating \citep[e.g.][]{young09}.  
Coordinate shifts relative to this wavelength should be applied based on the \verb+eis_ccd_offset+ routine in SSW.  
Interested users are encouraged to obtain the original EIS data from JSOC or other sources and prep it with the latest routines for further for analysis.

\subsection{DST/IBIS}
\begin{figure*}
    \centering
    \includegraphics[width=7in]{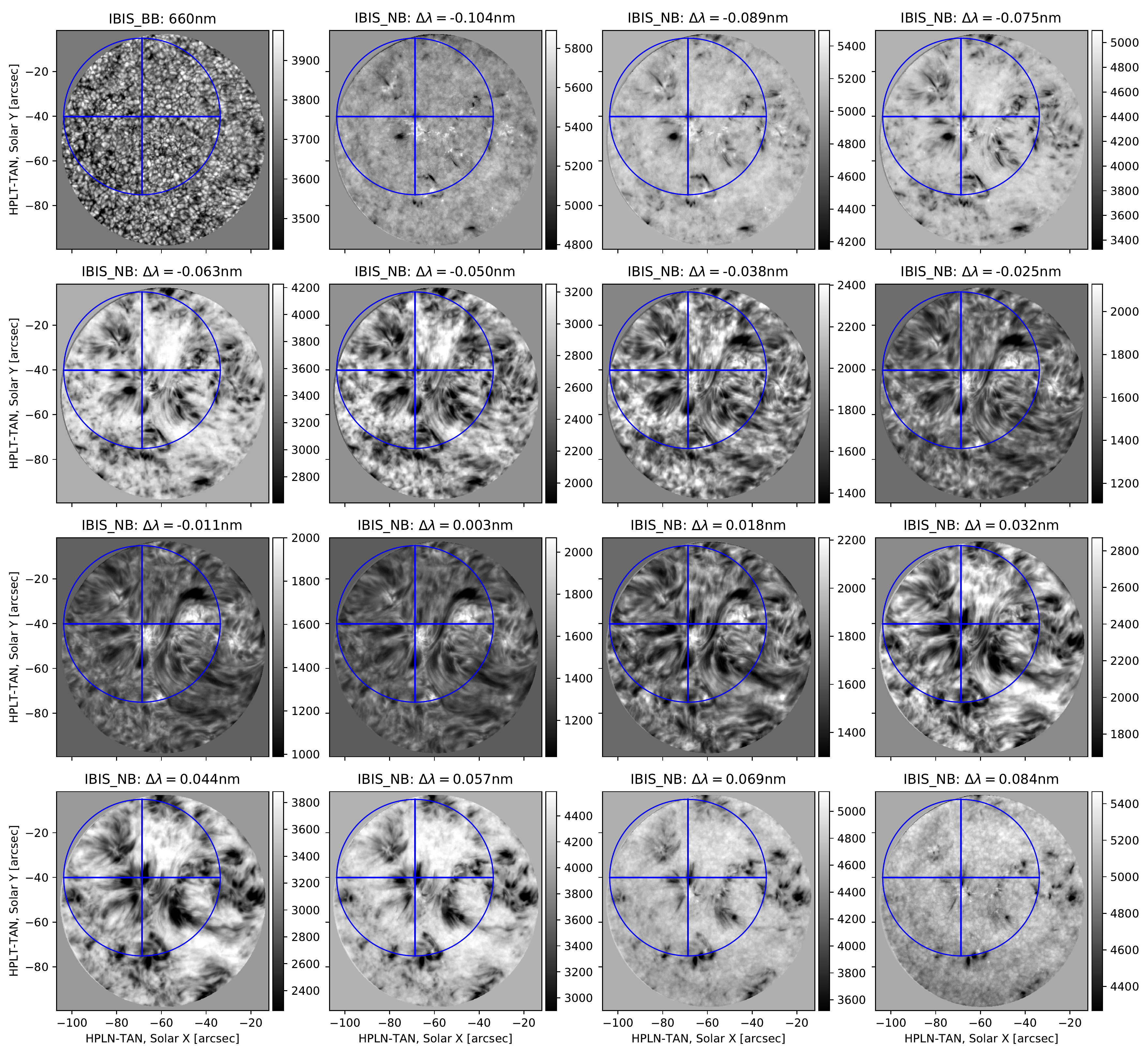}
    \caption{Co-aligned IBIS data for a partial scan through the \halpha{} line.  The scan began at 15:47:11.133 UT and lasted until 15:47:15.123, with approximately 0.165 s between each narrow band image. The broad band data is in the upper left while remaining panels show a subset of the narrow band images.  The blue circle and crosshair shows the ALMA field of view at 15:47:13 UT.}
    \label{fig:IBISgrid}
\end{figure*}

The IBIS broad band images were aligned to HMI continuum images in order to determine the final IBIS coordinates.
\figref{fig:IBISgrid} shows representative IBIS data, including a broad band image of the solar granulation pattern and a subset of the wavelengths from the narrow band data.  
The blue circle and cross shows the ALMA FOV at this time.

All co-alignment steps used the \code{chi2\_shift} cross correlation routine and then were independently verified using the \code{phase\_cross\_correlation} routine from the \code{skimage.registration} Python package.
The co-alignment process found a $1.08^\circ$ rotation of the IBIS data to align the pixel array with the solar north-south axis, while the coordinates of the center IBIS pixel were determined to be $(-76.041, -52.032)$ at time 2017-03-21 14:47:05.164.
The center-pixel coordinates for all other times were then calculated from that $(x,y,t)$ triplet using the \code{solar\_rotate\_coordinate} routine from SunPy 2.0.3.
The final offsets from the IBIS self-alignment described in \secref{sec:ibisselfalign} and the rotation to align pixel axes with solar X-Y axes were simultaneously applied to both the broad band and narrow band data using the \code{affine\_transform} interpolation routine from Sunpy. 

Each IBIS FITS file contains data from a single exposure time, e.g., a single broad band and narrow band image pair.
The final coordinates were saved in the WCS compliant CRVAL1 and CRVAL2 fields of the FITS header for each IBIS data file, which apply to both the broad band and narrow band images, as part of the primary FITS header.
The broad band and narrow band data are saved in the first and second Image FITS extensions, respectively, along with additional header fields pertinent to each.
Because of the limited FOV, disk-center location, and short duration $(\sim 2\unit{hr})$ of these data, we did not consider image distortion due to the curved surface of the Sun, as such corrections would enter at sub-pixel scales.\par

\subsection{ALMA}\label{sec:almaalign}
As explained by \citet{2019Molnar}, spatial variations in the Band 3 ALMA brightness temperature track variations in \halpha{} line width.
We therefore co-aligned the ALMA data to the IBIS-\halpha{} line widths calculated using the method described in Section \secref{sec:linewidth}.

The ALMA data were shifted and rotated to center the beam at the center of the pixel array and align solar north-south axis aligned with pixel columns.  
The FOV center coordinates were then varied such that levelsets of the co-temporal ALMA and \halpha{} data aligned at beginning of the ALMA data series at 2017-03-21 15:42:13 UT.  
As with the IBIS data, the WCS compliant coordinate values `CRVAL1' and `CRVAL2' at all other times were calculated by rotating the co-aligned center pixel coordinates from the initial time using the \code{solar\_rotate\_coordinate()} routine from \code{Sunpy}.

\section{Analysis}\label{sec:analysis}
\subsection{Target Overview}
To get a baseline qualitative understanding of how the active area evolved leading up to our primary observations we applied the segmentation and feature tracking algorithms described in \citet{2012Tarr} to the $\approx330\arcsec{}\times250\arcsec{}$ cutout of HMI data between 2017-03-19 00:00 and 2017-03-21 23:46 UT.  
This analysis revealed fairly standard network behavior.
Throughout the entire FOV, magnetic concentrations cyclically coalesce and fragment.
On short timescales (several hours) the motion of flux concentrations appears coherent, but on longer time scales ($\sim1$ day) movement appears random.   

However, the observed motions are not completely random.
Our observations are centered on the bipolar grouping, with positive magnetic fields to the west and negative to the east, shown in the right panel of Figure \ref{fig:SOTSP}.
Despite the continual fragmentation and coalescence, the bipole is present for the duration of the HMI data in the time period listed above and is associated with persistent, short, bright coronal loops in the EUV and X-Ray data.
As described below, it is likely that the bipole emerged as a cohesive unit that is well into the decay phase, but it may simply be a long-lasting random bipolar concentration. 
For ease of language we will refer to it as ``the bipole.''

Considered from a global perspective, the positive polarity is parasitic within a roughly circular medium-scale ($\sim250\arcsec{}$ diameter) negative polarity region. 
The medium-scale negative polarity region is itself surrounded by a larger-scale, predominantly positive polarity region that extends across both sides of the solar equator.
This magnetic configuration makes the bipole topologically isolated from other larger-scale features: magnetic domains are layered somewhat like shells in an onion and we are interested in the dynamics of the inner layers.
This larger-scale configuration will inform MHD simulations to be presented in future work.

The pattern of a medium-scale dominant negative polarity surrounded by a larger-scale positive polarity persisted for several solar rotations prior to our observations.
The bipole we are looking at is likely the decaying remains of AR 12639 which emerged around 2017-02-24 at $\approx$(10S, 30W)$^\circ$.

Returning to the local scale of the bipole, overall during our observations it appears to be decaying, both by apparent direct cancellation between the two polarities and by fragmentation and separation of each polarity individually.
The general decay of the bipole is complicated by two other factors: (1) nearby small-scale emergence, for example, around $(-475,-50)\arcsec{}$ at time 2017-03-19 13:45 UT, acts to replenish the decaying flux of each polarity, and (2) surrounding network concentrations that do not seem associated with emergence also random-walk into the bipolar region and replenish lost flux.

Finally, we note that immediately preceding our coordinated observations the parasitic positive polarity region of the bipole began an extended cycle of combined fragmentation and cancellation with the negative polarity of the bipole.
All of the processes described above at the photospheric level likely drive both wave activity and continual reconnection higher in the atmosphere, resulting in the transient brightenings we describe below.

\subsection{IBIS line widths}
\label{sec:linewidth}
\begin{figure}
    \centering
    \includegraphics[width=\linewidth]{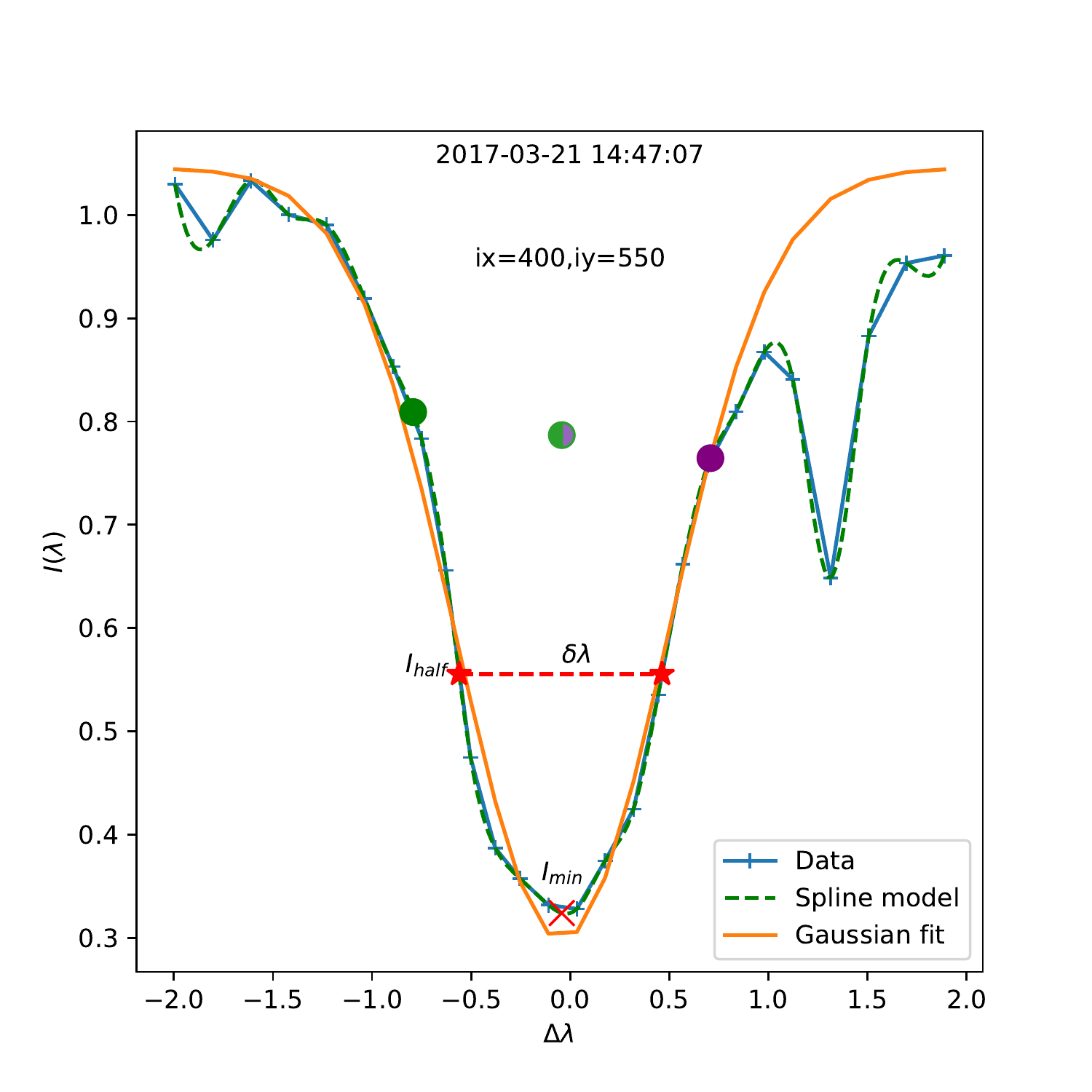}
    \caption{An example of the line fitting method, applied to the spectra at pixel $ix,iy=400,550$ at time 14:47:07 UT.  The blue line with `+' symbols is the observed normalized spectra.  The green dashed line is the spline interpolated model of the data.  The orange line shows the Gaussian fit to the spline model between $\approx-0.894$ \AA{} and $0.8369$ \AA{}.  \imin{} is marked by a red $\times$, the line intensities at $\linecenter\pm0.75$ \AA{} by the green and purple dots, their average by the multi-color dot, the width--defining intensity level by the red stars, and the resulting line width by the dashed red line.}
    \label{fig:linefitexample}
\end{figure}

We characterized the spectral profile of \halpha{} at every spatial and temporal point of the IBIS data series.
Figure \ref{fig:linefitexample} shows a typical example of the spectral data (green crosses), including the telluric oxygen line $\approx1.4$ \AA{} redward of the \halpha{} line center, pulled from a single spatio-temporal pixel.
Fitting the spectral line serves dual purposes.
First, as stated above in \secref{sec:almaalign} and reported on by \citet{2019Molnar}, the \halpha{} line width has proved useful for aligning the ALMA Band 3 data; we discuss this at length in the following subsections.
Second, the fitting produces several interesting plasma diagnostics of its own, including Doppler shifts and opacity fluctuations, which will be discussed in later publications.

Our spectral characterization followed the method of \citet{2009Cauzzi}.
\figref{fig:linefitexample} shows each step of the method applied to a single scan of the \halpha{} spectrum, pixel (400,550) at 14:47:07 UT.
For each spatial and temporal point we created a spline-interpolation model, $\hfit[\lambda]$ (green dashed line), of the normalized spectrum (blue solid line, with crosses)\footnote{The spectrum at each spatial pixel at a given time were normalized to the intensity of the first measured wavelength position averaged over the central $600\times600$ pixels.  That wavelength is $6560.832$ \AA{}, or $-1.9944$ \AA{} relative to the center of the prefilter bandpass.} using the \scipy{} \fnc{interpolate} package, based on FITPACK.  
Second, we determine the location of the line center by fitting a 3-parameter Gaussian absorption profile (orange curve) to the central 11 points in the line core, spanning -0.7505 to 0.5687 \AA{} from the center of the IBIS prefilter:
\begin{equation}
    G[\lambda; b, \linecenter, a] = b- \sqrt{\frac{\log2}{\pi a^2}}\exp\Bigl(-\frac{(\lambda - \linecenter)^2}{a^2\log2}\Bigr)
\end{equation}
where $b$ is the continuum level, $\linecenter$ is the line center, and $a$ is the half-width-at-half-max value.  
We only use the line center value $\lambda_0$ of this fit because the fit to the line center is fairly good even though the \halpha{} line profile is poorly represented by a Gaussian.  \par

We use the spline model and the fitted line-center to define the remaining characteristics of each spectral line.
Continuing with our example in \figref{fig:linefitexample}, the value of the spline model at line center defines the minimum line intensity, $\imin{}=\hfit[\linecenter{}]$, indicated by the red cross.  
Next, the line intensity at half the line depth is defined as the halfway point between the minimum intensity and average of the intensities at $\pm0.75$ \AA{} from \linecenter{}: 
\begin{equation}
    \ihalf = \frac{1}{2}\Bigl(\frac{\hfit[\linecenter-0.75\text{\AA{}}]+\hfit[\linecenter+0.75\text{\AA{}}]}{2}+\imin \Bigr).
\end{equation}
The intensities at $\pm0.75$ \AA{} are shown by the green and purple dots and their average by the central mixed green/purple dot.
The reference points at $\pm0.75$ \AA{} are chosen to avoid influence from the Telluric line at $+1.4$ \AA{} from line center.
Finally, the line width \linew{} is defined to be the difference between roots of the equation \mbox{$\hfit[\lambda]-\ihalf = 0$}, that is, the width of the line at intensity \ihalf.
Those points are indicated by red stars in the Figure, and the line width by the horizontal red dashed line.\par

\subsection[Line width vs brightness temperature]{Comparison of \halpha{} line width to ALMA brightness temperature}\label{sec:thalpha}

\begin{figure*}
    \centering
    \includegraphics[width=\linewidth]{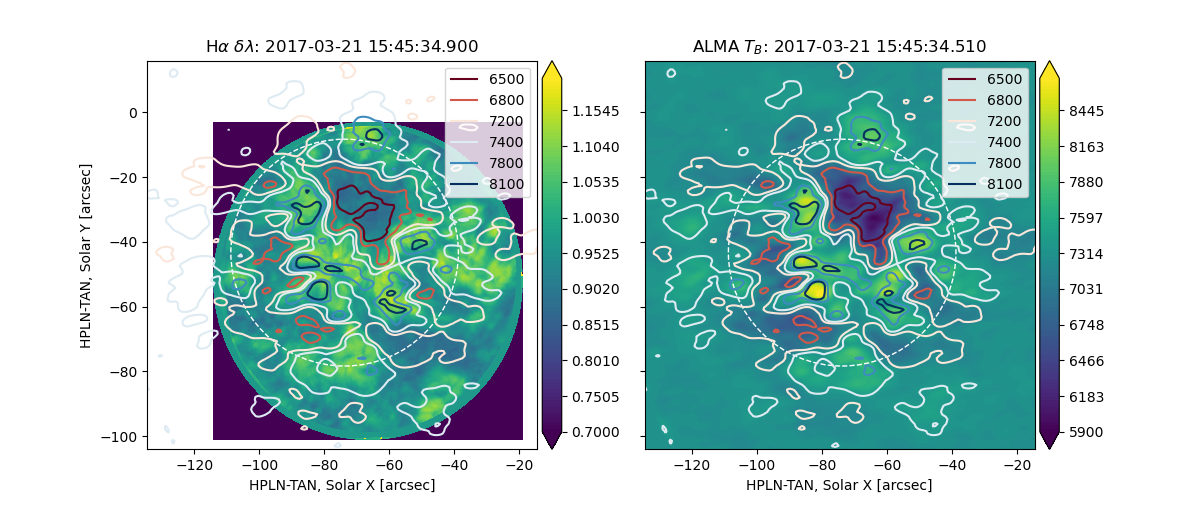}
    \includegraphics[width=\linewidth]{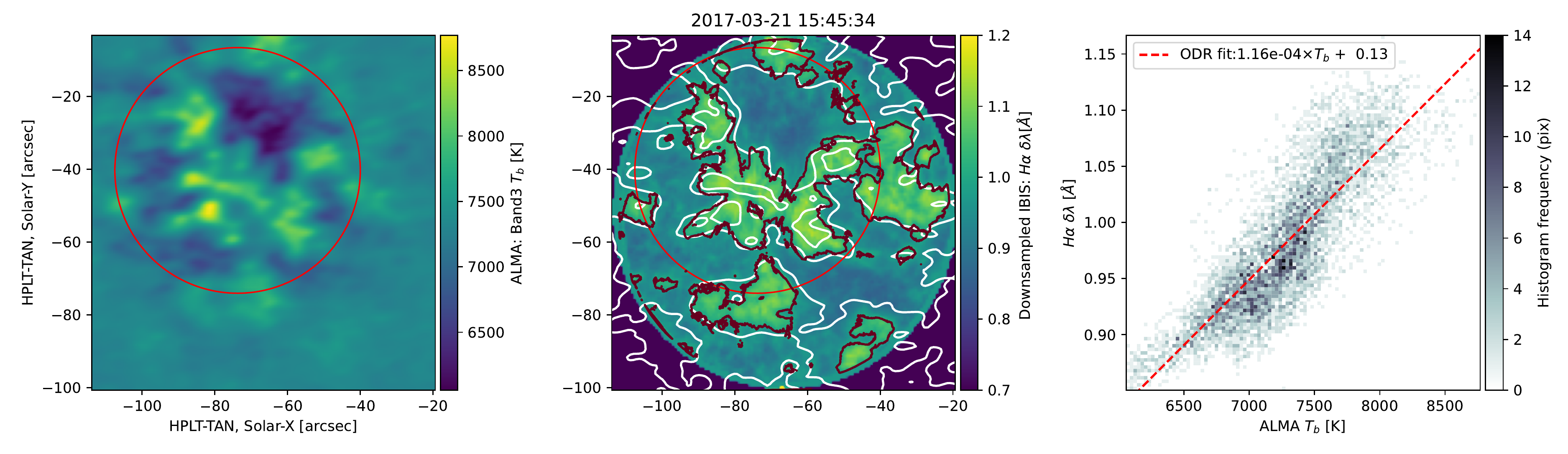}
    \caption{\emph{Top:} \halpha{} line-core widths from IBIS (left) and the ALMA Band 3 brightness temperature (right) at time 2017-03-21 15:45:34.  Contours of the ALMA brightness temperature are overlaid on both panels.  The white dashed circle shows the effective extent of ALMA's FOV. \emph{Bottom:} ALMA brightness temperature (left), \halpha{} line width down-sampled to the ALMA resolution (middle) with contours at $\delta\lambda=1$ \AA{} (red) and $T_B=7300K$ (white); and a 2D histogram and linear fit to the relation between line width and brightness temperature within the red circle of the preceding images (right).}
    \label{fig:almaibis}
\end{figure*}

The line-fitting procedure described in the previous section was applied to every spatial and temporal location in the IBIS data series, but for the rest of this section we focus on the single time at 15:45 UT shown in \figref{fig:almaibis}.
The trends we discuss below do hold throughout the $\sim1$ hour of co-temporal data, but a detailed analysis of the joint dynamics of the two series is outside the scope of the present work.

The top left panel \figref{fig:almaibis} shows a spatial map of the \halpha{} line width at 15:45 UT and the top right panel shows the ALMA brightness temperature map from the same time.
The dashed circles indicate the approximate extent of the reconstructed ALMA field of view of $\sim60\arcsec{}$.
The contours in both panels are of the ALMA brightness temperature and clearly show the good correspondence between the \halpha{} line width and the ALMA brightness temperature.
It is also clear that the correspondence extends somewhat outside of ALMA's effective FOV before fading into the background noise level; this is as expected.
We have therefore verified the qualitative correlation between \halpha{} line widths and ALMA Band 3 brightness temperature, as reported by \citet{2019Molnar}.

To quantify the relation between the ALMA Band 3 brightness temperature and \halpha{} line width we spatially down-sample the line width map to match the ALMA resolution and compare the results point-by-point.
The bottom row of \figref{fig:almaibis} shows the ALMA brightness temperature without contours (left); the down-sampled IBIS line widths (middle) overplotted with contours of the 1 \AA{} \halpha{} line width (dark red) and 7300 K ALMA $T_B$ level (white); and a 2D histogram (right) of the two previous plots using the cospatial data outlined by the red circular region of radius $33.75\arcsec$ centered on the ALMA beam center.
A linear fit to the 2D distribution using the orthogonal distance regression method\footnote{See the \fnc{scipy.odr} package.} is given in the figure legend and shown as the dashed red line.

Comparing the bottom right panel of \figref{fig:almaibis} to Figure 4 of \citet{2019Molnar} shows that we find roughly double the $\linew{}/T_b$ slope in this region of quiet Sun compared to their region of active region plage: $1.15\times10^{-4} \linew{}/T_b$ for our study compared to $6.12\times10^{-5}$ for theirs.
The cause of this discrepancy is currently unclear, but could be due to a number of factors.
First, the two data series sample fairly different physical conditions on the Sun.  Our ALMA observations are of quiet Sun to (minimally) enhanced-network conditions and span a much smaller range of temperatures compared to those found in \citet{2019Molnar}'s active region plage case. Second, our method for fitting the \halpha{} spectral line and defining the line width differs slightly from that of \citet{2019Molnar}: we used a narrower region of the spectrum to avoid the influence of the barely resolved (in our data) Telluric line at $+1.4$ \AA{} from line center.
These two issues confound a direct comparison between the two studies.
At the same time, the fit to our data does appear to adhere more close to the trend of their B, D, and F model atmospheres, taken from \citet{Fontenla:2011}.  These models correspond to the range between quiet Sun and enhanced network conditions, which again are more appropriate for our data than for those in \citet{2019Molnar}.

Regardless, both studies clearly demonstrate a strong relation between the width of the \halpha{} line and the ALMA Band 3 brightness temperature.  
How precisely that relation depends on the observed region, and how it might change under dynamic evolution, and what care needs to be taken when defining the line width, remains to be determined.

\citet{2019Molnar} synthesized the radiative emission from a range of model atmospheres, representative of quiet Sun to strong plage conditions, using the RH code \citep{2001Uitenbroek}.  
Those results suggest that the broadening of \halpha{} as the chromospheric temperature rises is primarily an opacity effect due to an enhanced number density of H atoms in the $n=2$ quantum state, as opposed to thermal broadening.  
They explained this behavior, extensively referencing \citet{2012Leenaarts}, as follows:
The \halpha{} source function is nearly uniform throughout the chromosphere.  
This feature gives \halpha{} its characteristic flat bottom, with fairly uniform intensity in wavelength moving away from line center, until a wavelength is reached at which the source function becomes sensitive to the photosphere, giving rise to the steep line wings.  
As the formation height of the line core increases, the transition to photospheric-dominated emission occurs further in wavelength from the line center.  

Now consider how this behavior changes as one varies the temperature of an atmospheric model.  
Moving from cooler to hotter models, two important things happen simultaneously \citep[see][Fig.~5]{2019Molnar}: (1) the $\tau=1$ surfaces for the \halpha{} line core and the ALMA Band 3 emission begin to coincide; and (2) the contribution functions for each line become more spatially localized higher in the atmosphere.
Thus, in the hotter models, which produce the largest \halpha{} line widths and highest ALMA Band 3 brightness temperatures, essentially all the \halpha{} absorption occurs in a thin range of heights base of the transition region and coincides with the ALMA Band 3 millimeter emission.  
However, the above \halpha{} formation behavior, and the resulting correspondence between \halpha{} and ALMA millimeter emission, eventually breaks down for the hottest models that are most similar to active regions, when the \halpha{} source function is no longer flat.  
We do not expect this break down to occur for the region of enhanced network we have targeted in this data set, with the result that we find good correspondence between the \halpha{} line width and ALMA brightness temperature throughout our observations.

The main result of this analysis is that we have qualitatively verified the relation found by \citet{2019Molnar} but our quantitative analysis produces a different value for $\linew{}/T_b$ slope.
Our observations include regions of only slightly enhanced network as opposed to theirs of strong, active region plage.
If the radiative modeling described above is correct then we would expect stronger correlation of temporal dynamics in the hotter-ALMA/broader-\halpha{} than in the cooler-ALMA/narrower-\halpha{} regions, but precisely how the dynamics would diverge, and where a break (if one exists) would occur is currently unknown.
Details of that temporal analysis will be presented in a forthcoming paper (Tarr et al., in prep.).

\subsection{Transient brightening detections}
While faint and relatively cool, this data set is quite dynamic in all observed wavelengths. The dynamics appear more impulsive in the corona, and are more readily detected as transient brightenings. Over 57 individual brightenings were identified in the XRT data series using the automatic detection techniques of \citet{2014KobelskiEA_XRTTBs}. Most of the detections were found within overlapping fields of view (spatially and temporally) of the other instruments. A few brightenings overlap with EIS and IRIS slits, but we have not yet interpreted these specific features. 

\begin{figure}
    \centering
   \includegraphics[width=0.9\linewidth]{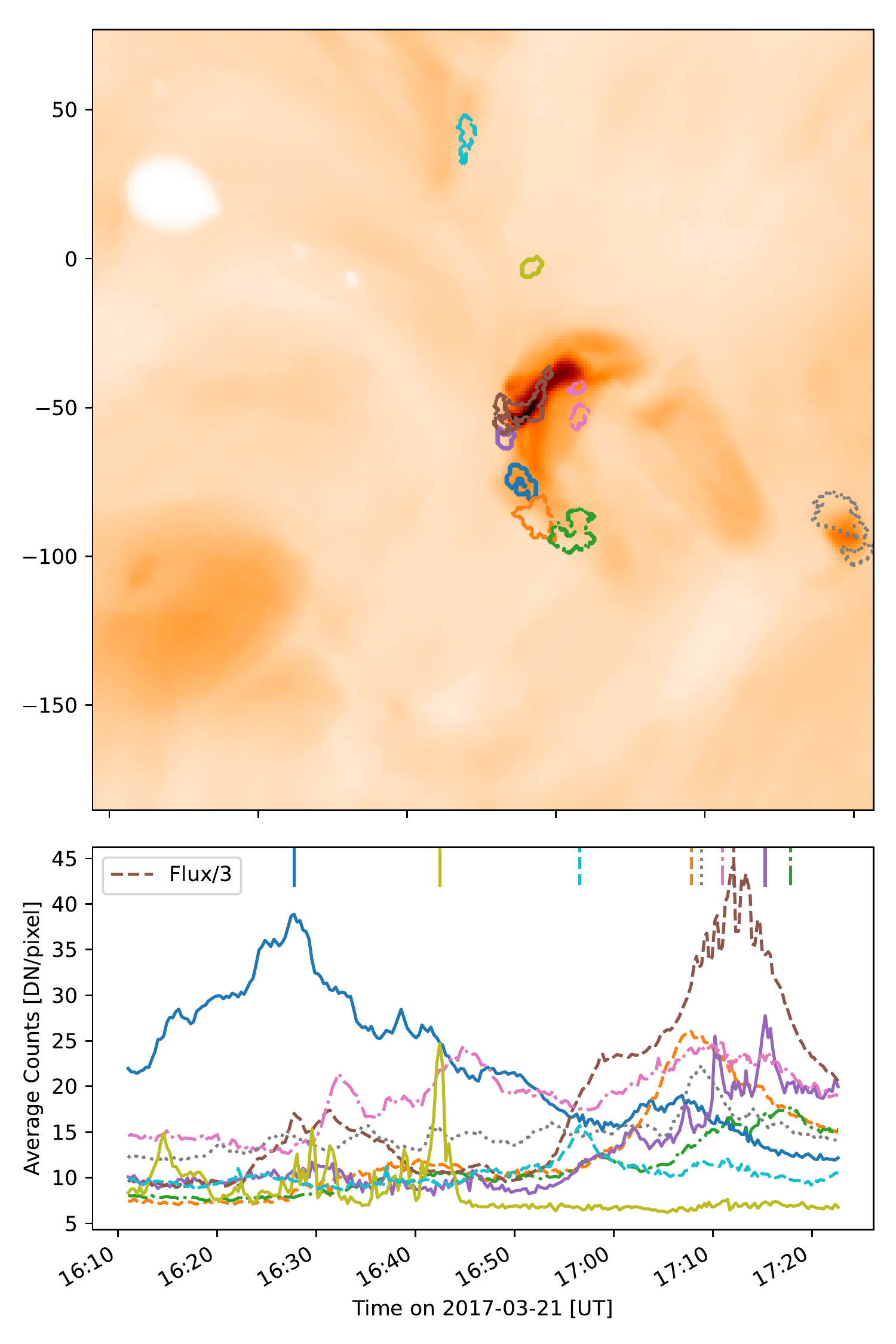}
    \caption{Upper panel: Median XRT Al Poly image from the time range from 16:11 to 17:22 (a continuous block of XRT observations as shown in Figure~\ref{fig:coordination}). The color scale has been reversed to show details. Overlaid on the image are 8 regions where transient brightenings were detected. Spatially averaged light curves for these regions are shown in the lower panel. Marks at the top denote the timings of the light curve peak for each region. The light curve for the main loop structure has been scaled down by a factor of 3 to better match the vertical scale of the other light curves.}
    \label{fig:xrtlcs}
\end{figure}

The brightenings could be consolidated to 8 distinct regions, as shown in Figure~\ref{fig:xrtlcs}. These regions are all made of individually detected brightenings, indicating significant substructure and dynamics. 

Note the progression of brightenings for the blue, orange, and green regions near the southern footpoint. These three regions appear above the bipolar region visible in SOT and HMI magnetograms south of the main bipole, and also coincide with the ends of loops structures observed in coronal EUV channels (see \figref{fig:coordination2}). We thus interpret the brightenings to be footpoints of a loop structure south of the main region. 
All three regions brighten twice, early around 16:25-16:30 UT, and later around 17:10 UT. 
However, the inner most region (blue) lights up most during the earlier event (around 16:15 UT), and shows only weak brightening at the later time. The other two footpoint regions show weak brightenings early, but more significant brightenings later. 
The progression shows brightening further from the main region, suggesting the successive brightening of loops between the upper photospheric flux concentrations and more southerly concentrations (as shown in the the upper left panel of Figure~\ref{fig:coordination2}).
In the next section we focus on a single dynamic event which was readily observed in most of the imaging data.  

\subsection{Transient brightening at 16:10-16:30 UT}
\begin{figure*}
    \centering
    \includegraphics[width=\linewidth]{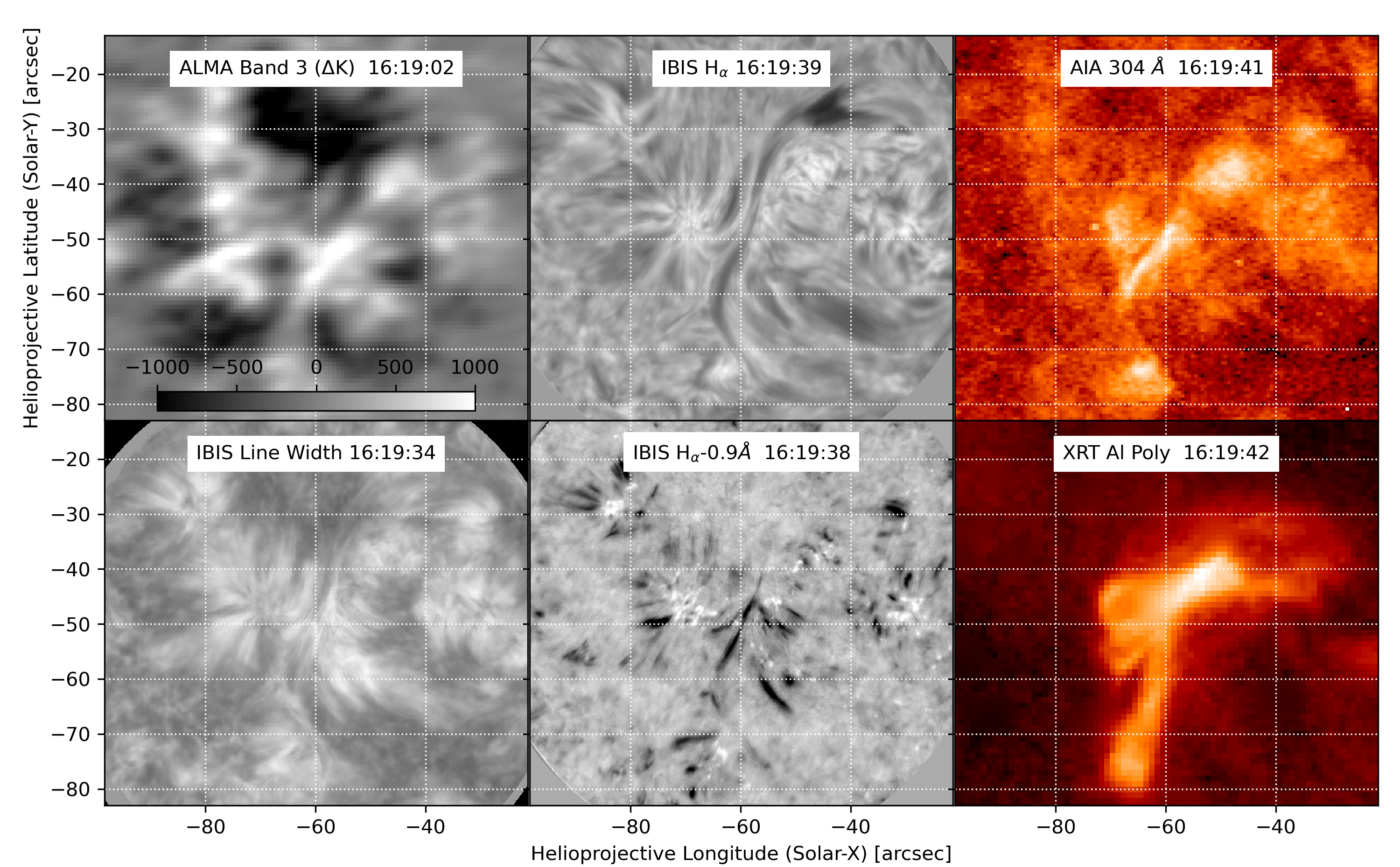}
    \caption{Subimage of the brightening event. This particular time step was chosen to showcase the filament feature in the ALMA (upper left) near (-60\arcsec, -55\arcsec). This feature is coincident with the blue shifted, filamentary absorption feature visible in the \halpha{} line wing (top center). This blue shifted feature persists for approximately 10 minutes, and as it fades new loop structures appear in the upper chromospheric and coronal channels, displayed in the bottom half of the panel. An animation of this figure is available online.}
    \label{fig:subset}
\end{figure*}

\begin{figure*}
    \centering
    \includegraphics[width=0.8\linewidth]{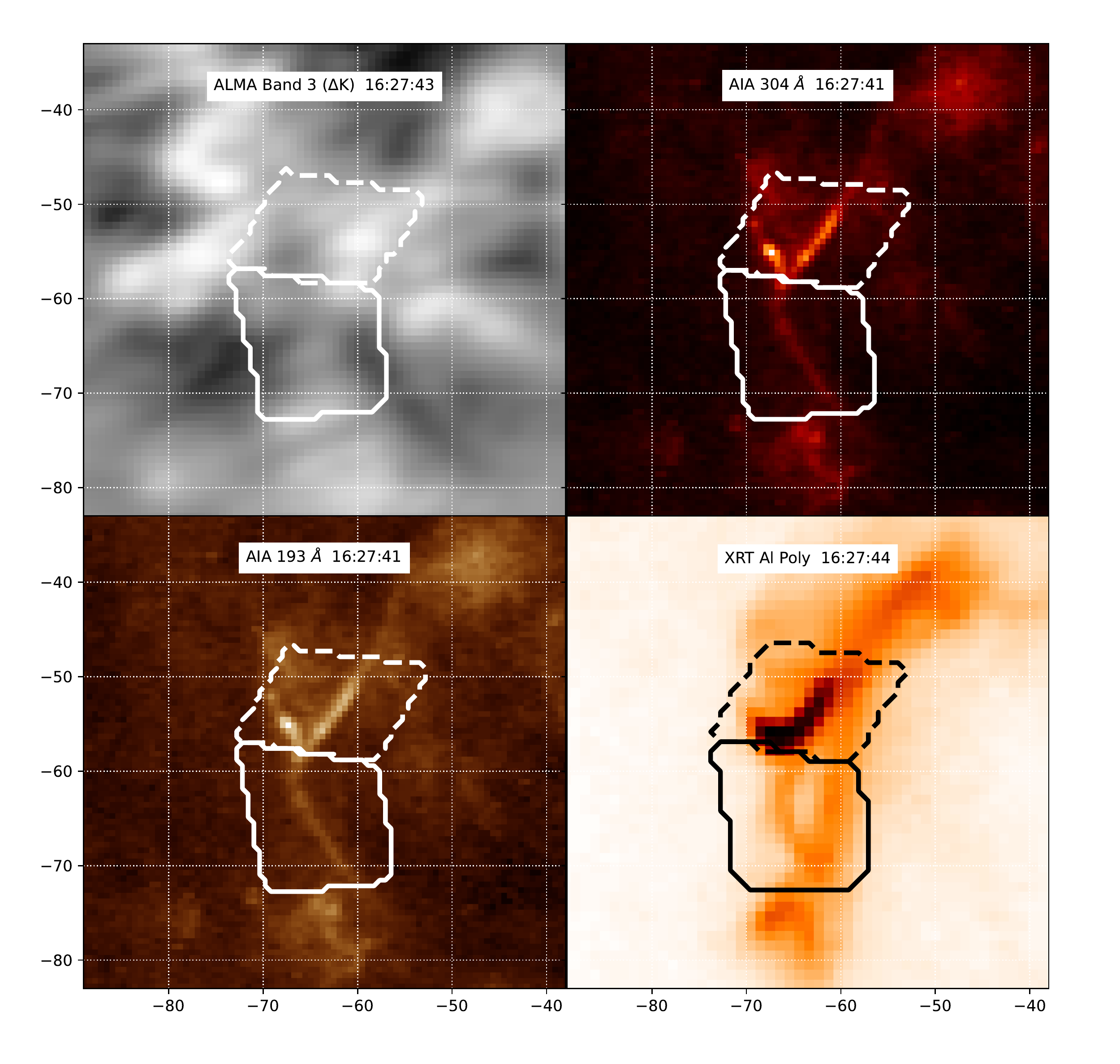}
    \caption{Sub-regions  shown in Figure~\ref{fig:lightcurves}. The bounding boxes were chosen from the AIA 304 \AA image, and made wide enough to cover any lingering instrumental drifts, loop motion, and variations between wavelengths. We have divided it into two distinct sub-regions. The upper region bounds the bright feature, and the lower box the `wishbone'. The reference time is chosen to highlight the coronal wishbone shown in the lower box. This timing occurs after the peak ALMA brightening (shown in Figure~\ref{fig:subset}). The XRT image has been reverse scaled to improve visibility. The brightening follows much of the same shape in all upper chromospheric and coronal wavelengths.}
    \label{fig:wishbonezoom}
\end{figure*}

We find evidence of a transient event in multiple data channels that appears to show energy transfer between the chromosphere to the corona and back along two different paths, between 16:10 and 16:30 UT.
\figref{fig:coordination2} shows nine of our data channels near the end of the event, with the timings of Figures~\ref{fig:subset} and~\ref{fig:wishbonezoom} chosen to showcase the regions at different points in the morphological development.
Animations of Figures \ref{fig:coordination2} and \ref{fig:subset} clearly show the dynamic evolution in multiple data series.

The event starts as an elongated strong dimming the \halpha{} blue wing, see especially the $-0.9$ \AA{} channel, as showcased in Figure~\ref{fig:subset}.
The dimming progresses roughly from SE to NW along the filament channel between the polarities of the bipole.
The ALMA data show cospatial enhanced temperatures during this time. 
The brightening is then observed in the coronal EUV and X-ray data series and follows the same progression from SE to NW along the filament channel, just as in the chromospheric IBIS and ALMA data series.

After brightening in ALMA, the dynamic event becomes visible in the hotter temperature and typically higher elevation data channels, especially AIA 304 \AA{} and 193 \AA{}, followed by XRT. 
This brightening is visible in the main cusp-shaped loop structure in the northern section of the region, but also manifests a small scale sympathetic event in the form of a wishbone-like loop feature directly to the south as shown in Figure~\ref{fig:wishbonezoom}. 
The wishbone feature  extends southward from the negative concentration of the bipole towards a small flux concentration near the southern edge of the ALMA field of view.
We observe a co-spatial and -temporal dimming in the \halpha{} blue wing and enhanced temperatures in ALMA at the southern terminus of the secondary loop. 

\begin{figure}
    \centering
    \includegraphics[width=\linewidth]{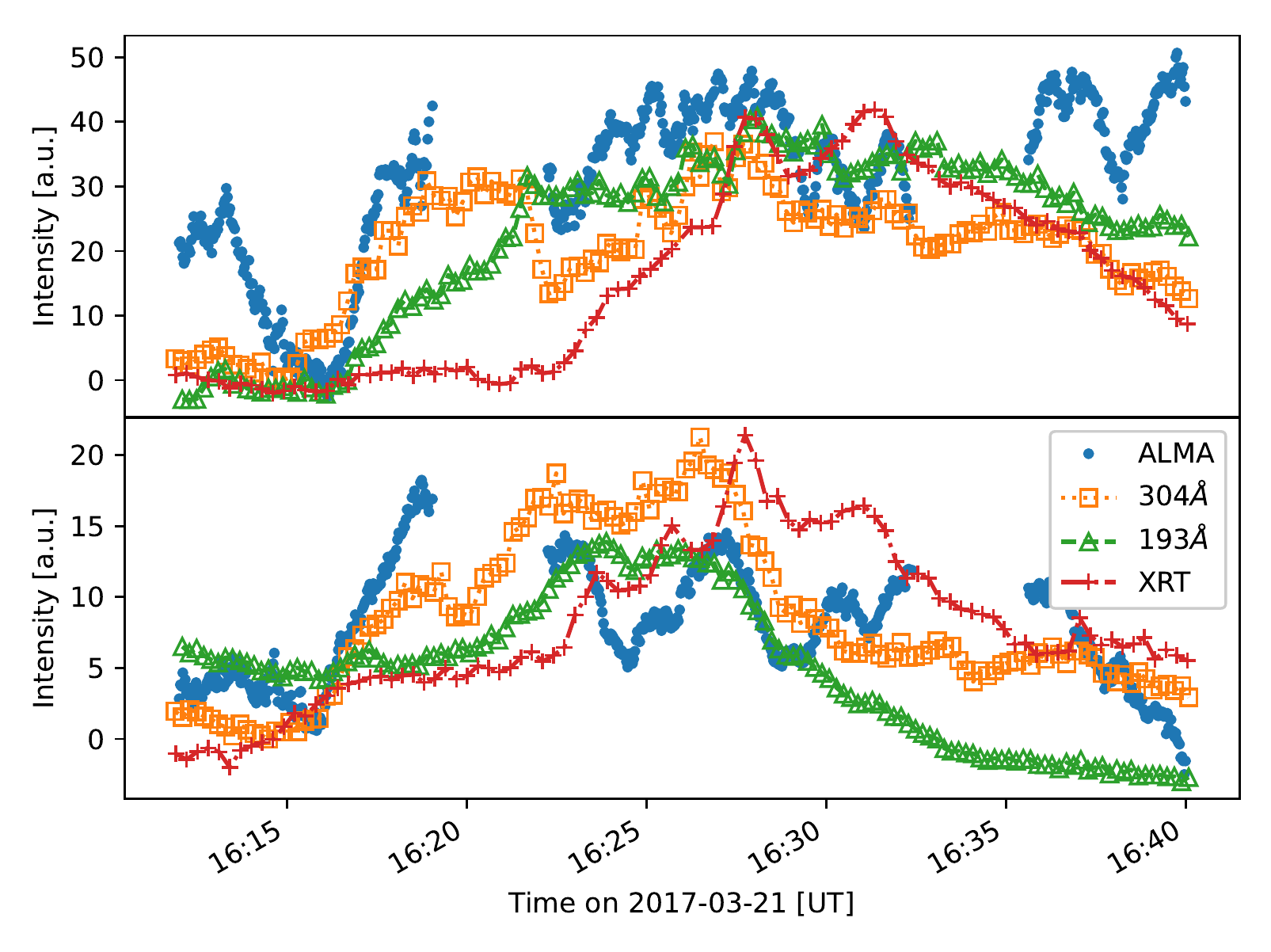}
    \caption{Light curves from upper and lower sub-regions shown in Figure~\ref{fig:wishbonezoom}. The brightening appears first in the low chromosphere as shown with ALMA. The EUV data then brightens next, with the AIA 304 \AA\ brightening slightly before the AIA 193 \AA, though these brightenings follow much of the same shape.}
    \label{fig:lightcurves}
\end{figure}

Figure~\ref{fig:lightcurves} plots light curves for each of the channels displayed in Figure~\ref{fig:wishbonezoom}.  Each curve in the upper (lower) panel is the average of the dashed (solid) box from Figure~\ref{fig:wishbonezoom}.
Note that these light curves have been scaled to fit on the same plot; the ALMA temperatures range from 7526 K to 7791 K on average in the upper region and 7064 K to 7271 K on average in the lower region.

The light curves clearly show the brightening starts in the low chromosphere with ALMA and is followed by the hotter channels.
The average temperature from the ALMA data within the upper box shows a brief dimming followed by a heating trend. 
This dimming appears to be from the region surrounding the filament, as the filament does not itself show notable dimming. 
The 193 \AA\ bandpass includes \ion{Fe}{12} and \ion{Fe}{24} lines which form in the corona and hot plasma at log(T)=6.2 and 7.3 respectively, as described in \citet{2012Boerner} and \citet{2014Boerner}. 
The closeness with which the AIA light curves track each other suggests an initial brightening in the lower temperature component of 193 \AA{} followed by brightening due to plasma heating the hotter component \citep[see Figure~11 of][]{2012Boerner}.

Finally, we observe brightening in XRT, which represents the hottest thermal features in the data set.
This is consistent with direct heating of the loop, but is in contrast to some observations of active region transients, including those involving ALMA such as \citet{2021Shimizu}. The difference between these results could be due to many factors, and highlight the need for more observational to understand these events.

\section{Conclusions}\label{sec:conclusion}
We have presented an extensive coordinated data set that includes data series from multiple facilities and, in some cases, multiple instruments from each facility.
We describe the calibration and self-alignment of each data series and the co-alignment of all data series within the data set.
The fully calibrated and aligned data are publicly available at \url{https://share.nso.edu/shared/dkist/ltarr/kolsch/}.
The coordination was keyed to our ALMA Cycle 4 observations which targeted a bipolar region of enhanced network for approximately one hour starting at 14:42 UT on 2017-03-21.
The target was situated close to disk center, approximately $(-70\arcsec{},-50\arcsec{})$ at the center of the ALMA data series. 
The other data series cover a variety of fields-of-view, start times, and durations.

We found that the spatial areas of broadest line width of the \halpha{} spectral line as observed by IBIS were cospatial with the hottest regions as measured by ALMA Band 3 $(\gtrsim 7500\unit{K})$.
This correspondence held for the entire duration of the ALMA observations.
This result extends the findings of \citet{2019Molnar}, who analyzed an area of active region plage, to quiescent solar regions.
However, the linear relation found in the two regimes differs by roughly a factor of two, and it is currently unknown if this is due to a difference in measurement method of the \halpha{} line width, properties of the targeted region on the Sun, or some combination thereof.
A future work will discuss the statistics of temporal dynamics between the two data series.

Preliminary analysis found multiple transient brightenings throughout the data set, some of which span multiple data series.
We highlight one particularly well observed example lasting approximately 20 minutes starting at 16:10 UT.
Light curves of the event show a clear transition from lower to higher temperature data series, starting in the chromospheric ALMA data ($\sim7000\unit{K}$) and progressing up to our hottest observed thermal data series in \textsl{Hinode}/XRT ($\sim3\unit{MK}$).
Spatially, the event shows a propagation first along a filamentary feature above the central polarity inversion line of the bipole and then a secondary Y-shaped extension to another network concentration approximately $15\arcsec$ to the south. 
This observation is somewhat different than the results of \citet{2021Shimizu}, which primarily found the ALMA observations to showcase chromospheric footpoint heating of a small flare.

What this data demonstrates is that these small scale events with large wavelength coverage are interesting, and cover some of the same dynamic features as seen in larger events. We hope the community is able to further utilize this small but robust data set.

\acknowledgments
The authors gratefully acknowledge the support by NASA grant 80NSSC19K0118.
This paper makes use of the following ALMA data: ADS/JAO.ALMA$\#$2016.1.00788.S. ALMA is a partnership of ESO (representing its member states), NSF (USA) and NINS (Japan), together with NRC (Canada), MOST and ASIAA (Taiwan), and KASI (Republic of Korea), in cooperation with the Republic of Chile. The Joint ALMA Observatory is operated by ESO, AUI/NRAO and NAOJ.

Data in this publication were obtained with the Dunn Solar Telescope of the National Solar Observatory, which is operated by the Association of Universities for Research in Astronomy, Inc., under cooperative agreement with the National Science Foundation.  
The authors thank K.~Reardon for help calibrating the DST/IBIS data.

IRIS is a NASA small explorer mission developed and operated by LMSAL with mission operations executed at NASA Ames Research center and major contributions to downlink communications funded by ESA and the Norwegian Space Center.

{\it Hinode} is a Japanese mission developed and launched by ISAS/JAXA, collaborating with NAOJ as a domestic partner, NASA and STFC (UK) as international partners. Scientific operation of the {\it Hinode} mission is conducted by the {\it Hinode} science team organized at ISAS/JAXA. This team mainly consists of scientists from institutes in the partner countries. Support for the post-launch operation is provided by JAXA and NAOJ(Japan), STFC (U.K.), NASA, ESA, and NSC (Norway).

This research has made use of NASA’s Astrophysics Data System Bibliographic Services.

\vspace{5mm}
\facilities{ALMA, Dunn(FIRS, IBIS), IRIS, SDO(AIA, HMI), Hinode(EIS, SOT, XRT)}

\software{\code{astropy} \citep{2013astropy,2018astropy}, \code{sunpy} \citep{2020sunpy}, \code{numpy} \citep{2020Harris}, \code{sswidl} \citep{1998FreelandHandy_SSW}}

\bibliographystyle{yahapj}
\bibliography{references}

\end{document}